\documentclass[12pt,preprint]{aastex}

\slugcomment{Accepted for publication in the Astrophysical Journal}

\shorttitle{Mapping Directly Imaged Giant Exoplanets}
\shortauthors{Kostov and Apai}

\begin{document}

\title{Mapping Directly Imaged Giant Exoplanets \\
   }

\author{Veselin B. Kostov\altaffilmark{1}}
\affil{Department of Physics and Astronomy, Johns Hopkins University, 366 Bloomberg Center, 3400 N. Charles Street, Baltimore, MD 21218}

\and

\author{D\'aniel Apai}
\affil{Department of Astronomy, The University of Arizona, 933 N. Cherry Avenue, Tucson, AZ 85718; Department of Planetary Sciences, The University of Arizona; Space Telescope Science Institute, Baltimore, MD 21218}

\altaffiltext{1}{To whom correspondence should be addressed. Email: vkostov@pha.jhu.edu}

\begin{abstract}

With the increasing number of directly imaged giant exoplanets the current atmosphere models are often not capable of fully explaining the spectra and luminosity of the sources. A particularly challenging component of the atmosphere models is the formation and properties of condensate cloud layers, which fundamentally impact the energetics, opacity, and evolution of the planets.

Here we present a suite of techniques that can be used to estimate the level of rotational modulations these planets may show. We propose that the time--resolved observations of such periodic photometric and spectroscopic variations of extrasolar planets due to their rotation can be used as a powerful tool to probe the heterogeneity of their optical surfaces. In this paper we develop simulations to explore  the capabilities of current and next--generation ground-- and space--based instruments for this technique. We address and discuss the following questions: a) what planet properties can be deduced from the light curve and/or spectra, and in particular can we determine rotation periods, spot--coverage, spot colors, spot spectra; b) what is the optimal configuration of instrument/wavelength/temporal sampling required for these measurements; and, c) can principal component analysis be used to invert the light curve and deduce the surface map of the planet. 

Our simulations describe the expected spectral differences between homogeneous (clear or cloudy) and patchy atmospheres, outline the significance of the dominant absorption features of H$_{2}$O, CH$_{4}$, and CO and provide a method to distinguish these two types of atmospheres. Assuming surfaces with and without clouds for most currently imaged planets the current models predict the largest variations in the J--band. Simulated photometry from current and future instruments is used to estimate the level of detectable photometric variations. We conclude that future instruments will be able to recover not only the rotation periods, cloud cover, cloud colors and spectra but even cloud evolution. We also show that a longitudinal map of the planet's atmosphere can be deduced from its disk--integrated light curves.

\end{abstract}

\keywords{giant planets: general --- direct imaging: surface features, photometric and spectroscopic variability, lightcurve inversion}

\section{Introduction}

Clouds play a fundamental but complex role in the atmospheres of brown dwarfs and exoplanets: describing their vertical and horizontal distributions, composition, formation and evolution are the outstanding challenges faced by models of ultracool atmospheres. Refractory minerals, such as Ca-Al-oxides, metallic iron and silicates, condense in the temperature range $\sim$1,300--1,900 K, forming clouds that dominate the atmospheres of L--type dwarfs and exoplanets of similar temperatures \citep[e.g.][]{burrows09, fortney08}. These silicate clouds are responsible for their very red near-infrared colors \citep[e.g.][]{burrows09,burgasser09,marley10}. Silicate grains, for example, have already been observed with Spitzer \citep{cushing06}, supporting this assumption.  In contrast, the spectrum of cooler T--dwarfs is markedly different --- it is characterized by blue near-infrared colors and is dominated by the absorption of stable gas--phase H$_{2}$O and CH$_{4}$. The spectra of T dwarfs are explained by clear, i.e. non-cloudy, models. The depletion of refractory elements at the L/T--transition regime (at temperatures of $\sim$1,000 to 1,200~K), caused by their "rain out" to deeper, hotter layers \citep{burrows09,burgasser09,allard10} results in a strong change in the spectrum which, combined with the cloud dispersal assumption, is the proposed mechanism for the onset of the T--dwarf regime.

Although the loss of cloud opacity at the L/T transition is qualitatively consistent with the drastic changes observed in the spectra, the process leading to the loss of clouds has not yet been identified. At least two different ideas have been explored. One plausible mechanism, supported by the work of \citet{knapp04} and \citet{tsuji03}, proposes a thinning (or even complete dissipation) of the global cloud cover, caused by the growth and subsequent sinking to deeper layers of the cloud particles. Another possibility, motivated in part by visible and near-infrared observations of Jupiter \citep{westphal74,orton96,dyudina01} and by near-infrared Cassini/VIMS observations of Saturn \citep{baines05}, proposes a sudden appearance of clear, optically thin "holes" in the global cloud deck \citep[][and others]{burgasser02,marley10}. In this scenario, in objects close to the L/T transition flux from deeper, hotter regions can escape through these holes. Therefore the atmosphere will appear patchy with bright, hot and deep regions next to cooler, optically thick higher--altitude ones. Both scenarios can produce similar near-infrared colors at the L/T transition by varying different parameters -- temperature, surface gravity, sedimentation efficiency and/or grain size for the former and cloud-cover and distribution for the latter \citep{saumon08}. The predicted differences are subtle which, when combined with the uncertainty in the measured temperature, currently makes it difficult to distinguish these models \citep{marley10}. For example, models that can explain the observed broadband photometry of L/T brown dwarfs do not guarantee a proper match for their near-infrared spectra \citep{burrows06}. This was clearly shown for the case of HR8799b \citep[e.g.][]{barman11a}, where the broad band photometry is well matched to an L6 dwarf while the near-IR spectrum is not. 

The differences found between the spectra of individual L/T transition objects and the model predictions, while difficult to explain, are not surprising as the current models do not account for the possibly complex three-dimensional atmospheres and large-scale atmospheric circulations, which also set the appearance of Jupiter. Therefore, explaining the atmospheric properties of the L/T dwarfs and giant exoplanets of similar temperatures, will likely require more than one-dimensional information. Understanding the spectra of these ultra cool atmospheres will require a physical model not only for the formation, evolution, and destruction of cloudy regions, but also for their longitudinal/latitudinal and vertical distributions. 

Not surprisingly, clouds also represent a key problem in the atmospheres of giant exoplanets of similar temperatures. Many of the recently discovered, directly imaged exoplanets fall in the temperature regime of the L/T transition: HR8799~bcde  with T$\sim$1,000K \citep[e.g.][]{marois10} and $\beta$ Pictoris with T$\sim$1,500K \citep{lagrange09, quanz10,skemer12}. These planets are cooler than Hot Jupiters ($\sim1,500 - 2,400$~K, \citealt[][]{seagerdeming10}), but much hotter than the effective temperature of Jupiter ($\sim$ 200K \citealt[][]{seiff98}). While directly--imaged planets may differ from brown dwarfs in bulk chemical composition, surface gravity and formation mechanism, their atmospheric physics is thought to be very similar.

The need for understanding cloud properties became even more pressing with the realization that many directly imaged exoplanets are significantly under-luminous in the near-infrared bands compared to field brown dwarfs and some state-of-the-art models: HR8799b (e.g. \citet{marois10, currie11,barman11a,skemer12}); 2M1207b \citep{mohanty07, patience10, skemer11}. Interestingly, $\beta$~Pic b -- a younger planet with estimated mass close to HR~8799b and 2M1207b -- appears to align well with model predictions \citep{lagrange10,quanz10,bonnefoy11}. In contrast, the prominent underluminosity of HR8799b and 2M1207b sparked intense theoretical work and most groups were led to propose cloud properties that differ significantly from those assumed for field brown dwarfs \citep{barman11a,barman11b,madhusudhan11,skemer12}. Several groups found best photometric matches from a combination of thick and thin clouds and therefore argued for patchy atmospheres in the HR 8799 planets (e.g. \citet{marois08, skemer12}). For a more detailed discussion on the latest results we refer the reader to Section \ref{State_of_the_art}. 

While it appears that directly-imaged planets are not a new class of objects but a natural continuation of the L-dwarfs sequence (rather than of the T-dwarfs), these developments emphasized the need for developing a more realistic model for cloud properties which, in turn, demand new, multi-dimensional data. Photometric variations due to rotating, spectrally heterogenous objects have been proposed as a probe of the cloud properties in brown dwarfs (e.g. \citet{bailer99, bailer01, burgasser02}). Similar observations of Earth from the EPOXI spacecraft were used to probe land mass and ocean distributions \citep{cowan09}. While the first searches for varying brown dwarfs produced several tentative detections, the past years brought the detection of periodic, rotational variations in L/T transition dwarfs \citep{artigau09,clarke08,radigan12}. For example, a newly discovered several L/T transition brown dwarf shows a peak-to-peak near-infrared variation as large as 27\% in the J--band \citep{radigan12, apai12}. With short typical rotation periods ($<$10 hours) these sources reveal patchy cloud covers on brown dwarfs. 

The best example of rotationally--induced photometric variability of a giant planet is, not surprisingly, in Jupiter. Using IRTF and HST mosaics \cite{gelino00} simulated the photometric variability of an ÓunresolvedÓ Jupiter due to its rotation. At 4.78 $\micron$ Jupiter indeed shows a very strong rotational modulation (up to 0.2 magnitudes), detectable at 0.41 $\micron$ as well on the level of 0.04 magnitudes. The culprit for these variations is the Great Red Spot which manifests itself as a large dark patch in the thermal infrared (where the directly imaged giant planets are bright). Thus, while not at the temperatures of the current census of directly imaged planets, Jupiter can be used as a reasonable starting point for studying their expected optical appearance. 

The need for characterizing cloud properties, combined with the exciting new results on brown dwarf variability, and the strong indications of patchy dust clouds in directly imaged planets motivate our study to propose variability of directly--imaged giant exoplanets as a means for characterizing their cloud covers. We present here a model for the atmospheric appearance of a directly--imaged giant planet that can be easily modified to represent any scenario: clouds on a clear atmosphere, clear "holes" on a cloudy atmosphere, a global clear or cloudy atmosphere with cold and/or hot spots. Cold spots could represent cloud thickness/structrure variations and heterogeneity \citep{radigan12} which would change the brightness of the respective patch so we included both possibilities.  The goal of this paper is to explore the observational signatures of such heterogenous atmospheres on directly--imaged giant planets. The choice of instrument, wavelength, and cadence of the observations are all target--dependent and not obvious. New instruments can be optimized for exploiting this technique, but this requires an understanding of the variations  expected from the rotating giant planet targets. We also use the simulated lightcurves to deduce the longitudinal distribution of the eigencolors using principal component analyses and to recreate the longitudinal spot patterns of the input map.

Here we provide a framework for identifying specific photometric and spectroscopic signatures expected of future directly imaged giant planets. The paper is organized as follows. In Section \ref{ModelDescription} we explain the details of our model. Our results are described in Section \ref{Results}. We present the deduced longitudinal map of the planet in Section \ref{Map}, discuss the key points in Section \ref{Discussion} and draw our conclusions in Section \ref{Conclusions}.  

\section{Model Description}
\label{ModelDescription}

We constructed a model to predict photometric and spectroscopic variations as a function of exoplanet rotation phase, wavelengths of the observations, and instrument/telescope. The key properties of the target are its effective temperature and inclination, and the temperature, spatial and size distribution of the spots. We use the contrast limits for the current and future instruments provided in the literature to simulate the relative photometric and spectroscopic accuracy. Our model first generates a 2D spot distribution for the exoplanet, determines the rotational modulations in the integrated lightcurve and spectrum as a function of wavelength, and then simulates the observations with the selected instrument. We explore the possibilities of this technique for giant planets to guide future observations and identify requirements for various instruments to inform their development. 

We will first review and discuss the cloud model, then describe the spectral libraries we use and discuss the simulated observations.

\subsection{Exoplanet Cloud Models}
\label{Section2.1}

We model the spot distribution of the exoplanet with a combination of an ambient/mean spectrum, calculated for the effective temperature of the planet, and a set of elliptical spots with different temperatures. The spots can have any size, shape, spatial distribution, temperature, surface gravity, metallicity and covering fraction. In the following, when we speak about the features of our models we will refer to "spots", but clouds will be used in the context of the astrophysical objects in ultracool atmospheres.

As a guide for spot size and shape distribution we use Jupiter, our best yet imperfect analogue. As will be shown, the general results of this paper do not depend strongly on this choice or the specific cloud shape and distribution. The cloud pattern of Jupiter is described by latitudinal bands, correlated to zonal circulation \citep{vasavada05}. The visual appearance of Jupiter is dominated by ammonia clouds located between 250 $mbar$ and 1 $bar$ (West et al. 2004). Above them are hazes which are observed in the near--UV, while deeper atmospheric layers are probed outside methane bands in the near--IR \citep{barrado09}. 

We set up our initial models to resemble the atmosphere of Jupiter, using elliptical spots with an aspect--ratio of 1.5, based on the Great Red Spot (GRS) -- with dimensions of 12,400 km and 19,800 km it had an aspect ration of 1.59 in 2006 \citep{rogers08}. To emphasize the effect such a spot can have on the visual appearance of a giant planet, we note the size the GRS had a 100 years ago when its longitudinal extent was about 45,000 km \citep{irwin03} and it covered about 3\% of the total surface area of Jupiter. The largest attainable size of turbulent eddies, defined as the Rhines length, is a function of the atmospheric wind speed and the gradient of the Coriolis force \citep{showman10} (Section 3.6, Equation 35), neither of which are constrained for giant planets. We note that the maximum size of spots present in planetary atmospheres is not necessary equivalent to the Rhines length, as the nature of such spots may be significantly more complicated than simple cyclonic eddies with different temperatures. It is not unreasonable to imagine a rather exotic situation where the vertical structure of the atmosphere is such that multiple, stratified layers of hazes and/or cloud layers with variable thickness alternate in such a way that they do or do not obscure deeper/hotter regions in the atmosphere. Nevertheless, the Rhines length provides a reasonable initial scale and we adopt it as the parameter setting the size of the largest spots.

For Jupiter, the Rhines scale is on the order of 10\% the planet's radius. Atmospheric models of Hot Jupiters, however, have suggested the presence of very high wind speeds on the order of 1--3 km $s^{-1}$, \citep{showman10} (Section 3.3, Table 1) and the possibility of much larger Rhines scale (comparable to the size of the planet). With a similar radius to Jupiter and assuming comparable rotation rates but faster winds (between Jupiter's and those for the Hot Jupiters), the Rhines length for directly--imaged giant planets will be larger than that for Jupiter. As a consequence the largest spots may be larger. Thus, to construct our atmospheric map we use a single giant spot covering a 5\% fraction of the total surface area and a set of additional, smaller spots, distributing their semi--major axes as a power-law of $D(N)=A \times 10^{-\alpha}$, where D is an array of N semi-major axes (one for each spot), A is a scale--factor (described below) and $\alpha$ is the power-law index. The total number of spots, the power--law index, the total spot--covering factor ($f_{c}$ hereafter) and the spot aspect-ratio are free parameters in the model. For illustrative purposes, here we use N = 20 spots with $\alpha = 2.0$ and $f_{c} = 10\%$, in such a way that the largest 5 spots correspond to 85\% of the "patchy" contribution. The choice of parameters is such that a single giant spot dominates most of the signal, a few smaller spots produce smaller, albeit still detectable signatures (as discussed below) while the rest of the spots are the tail of the power law and too small to be detectable. We use the scale-factor A to scale the sizes of the spots such that the total area covered by the sum of all spots is the predefined $f_{c}$, while keeping their relative sizes, as determined by the power-law size distribution, unchanged:

\begin{equation} A = \frac{4\pi f_{c}}{\sum_{i=1}^N s_{i}}
\end{equation}

where $s_{i}$ is the surface area of the $i_{th}$ spot. 

To calculate what fraction of the surface the different spot types cover during the rotation, we first randomly distribute the spots on a sphere then project the hemisphere visible for discrete rotational phases and finally measure the rotationally modulated fractional coverage for each spot. We use orthographic projection \citep{snyder87}, a technique that represents the actual appearance of a distant planet. It does not preserve the size or the shape of the surface features but as we are interested in the disk--integrated lightcurve and not in the best cartography of its surface, this transformation is well suited for our purposes. The projection is defined as \citep{snyder87}:

\begin{equation}
x = R \cos(\phi) \sin(\lambda - \lambda_{0})
\end{equation}
\begin{equation}
y = R[ cos(\phi_{1}) \sin(\phi) - \sin(\phi_{1}) \cos(\phi) \cos(\lambda - \lambda_{0})]
\end{equation}

where $x$ and $y$ are the cartesian coordinates on the projected 2--dimensional map, $\lambda$ and $\phi$ are longitude and latitude on the sphere, $(\lambda_{0}, \phi_{1})$ are longitude and latitude of the center point of the projection and $\emph{R}$ is the radius of the sphere, which is unity in our model.

An example of the two projected hemispheres of the planet with an inclination of $0^\circ$ is shown in Figure \ref{fig1}, where the different colors correspond to different spots, as described in the figure. From this projected map we extract the rotationally modulated fractional coverage of each surface element as a function of the planet's period, phase angle and inclination as described above. Figure \ref{fig2} shows the contributions of the largest spots to the visible hemisphere as a function of rotational phase. The giant spot covers 21.5\% of the projected visible hemisphere, while the next five spots by size cover between 2\% and 4\% each. We note that the relationship between the total surface fraction covered $f_{c}$ and the surface fraction of each spot is a function of the inclination of the planet and the size and shape of the spots.
 
In the next section we describe how these covering fractions are used to combine the spectra of all spots present on the hemisphere facing the observer for each rotation phase.

\subsection{Spectra and Spectral Libraries}
\label{Section2.2}

To each unique surface element ("spot") we assign a model spectrum from one of two different spectral libraries -- \citep{burrows06} (B06 hereafter) models or the AMES models of \citep{allard01} (A01 hereafter). The free parameters of these model libraries are temperature, metallicity, $log(g)$ and cloudy or clear atmosphere. We combine the cloudy and the clear models from the same library, keeping the surface gravity and the metallicity constant for a given object. We note that this step implicitly assumes that the pressure-temperature distribution of each column of gas is independent of that of neighboring columns, a good first-order assumption, which is not strictly correct  \citep[][]{marley10}. 

We explore objects with broad temperature range around the L/T transition (T $\sim$ 700 K to 1,400 K), representative of the giant self--luminous planets current and next--generation facilities are expected to directly image. Throughout the paper we use model spectra for solar metallicity and $log(g)=4.5$ for simplicity. Model spectra for different temperature regimes for A01 (Cond) and B06 (both Clear and Cloudy) are shown in Figure \ref{fig3}.  The two Clear models are quite similar, the main difference being the more detailed features of A01. We note that the Cloudy B06 are significantly different from the Clear B06 both in the strength and in the shape of the spectra in all three filters shown in the figure, a feature that will be discussed in more details in the discussion section. 

Next, the spectra of all spots present on the hemisphere facing the observer are weighted by their respective surface cover fraction and linearly combined for each rotation phase as:

\begin{equation}
F_{\lambda, tot} = \sum_{i=1}^{N}[\emph{F}_{\lambda, i}\times f_{c,i}]
\end{equation}

where $\emph{F}_{\lambda, i}$ and $f_{c,i}$ are the flux density and covering factor respectively of each surface element on the visible disk of the planet and \emph{N} is the number of different surfaces on the visible disk. The photometry is calculated from the flux density by first normalizing the spectrum by the width of the different filters we explore and then integrating.

\subsection{Simulated Observations}
\label{Section2.3}

\subsubsection{Targets and Instruments}
\label{Section2.3.1}

As a target we assume a star and giant exoplanet resembling HR8799~c (at a distance of 40 parsec), with a rotation period of 4 hours (planet's apparent brightness: J = 17.65~mag, H = 16.93 mag, Ks = 16.33 mag,  $T_{eff}=1,000$~K around an A5V star with J = 5.38 mag, H = 5.28 mag and Ks = 5.24 mag, \citep{marois08}). Correspondingly, the planet--to--star flux contrast ratio is $\sim1.2\times10^{-5}$ in J--band, $\sim2.2\times10^{-5}$ in H--band and $\sim3.7\times10^{-5}$ in Ks--band.

We model four different setups representative to the current and next-generation facilities shown in Table \ref{Table1}: an 8m--class telescope with Adaptive Optics (8m AO) representing VLT/NACO and Keck AO; an 8m--class telescope with an Extreme--AO (8m ExAO) representing VLT/SPHERE, Gemini/GPI, LBT/AO; a 30m--class telescope with Extreme AO (30m+ ExAO) representing Giant Magellan Telescope (GMT), Thirty Meter Telescope (TMT), Extremely Large Telescope (ELT)); and the James Webb Space Telescope (JWST). While these examples do not include all planned instruments, such as the Lyot 1800 project \citep{oppenheimer04} or ATLAST \citep{postman10}, they bracket the range of relevant instrument capabilities for the next two decades.

To estimate the achievable photometric accuracy of the different instruments we rely on the residual radial contrast curves provided by the instrument teams: VLT/NACO at 4~$\micron$ \citep{kasper07,kasper09}, VLT/SPHERE in J--band \citep{vigan10,mesa11}, Gemini/GPI in H--band \citep{macintosh08}, TMT/PFI in H--band \citep{macintosh06}, ELT/EPICS in J--band \citep{kasper08,kasper10a} and JWST/NIRCAM in K--band \citep{green05}. All contrast limits are for coronographic images. For VLT/NACO, VLT/SPHERE, Gemini/GPI and JWST/NIRCAM we place the planet at a separation of 1$\arcsec$, similar to the "c" planet in the HR8799bcde system \citep{marois08}. For the 30m--class telescopes we assume a separation of 0.2$\arcsec$, due to the smaller field-of-view of some of these instruments. The sensitivity limits for the four different instrument classes we explore are shown in Table \ref{table2}. 

\subsubsection{Synthetic Photometry and Spectroscopy}
\label{Section2.3.2}

To explore our ability to recover surface details of giant exoplanets we create a set of simulated observations. Rotationally--modulated lightcurves in J, H and Ks--bands produced as described in Section \ref{Section2.2} for an exoplanet with an atmospheric map from Figure \ref{fig1} are used to simulate the flux from the planet measured by a suite of current and future instruments. Throughout this work we use the VLT/NACO filters defined with central wavelengths and widths (in $\micron$) as follows: J--band (1.265 and 0.25 respectively); H--band (1.66 and 0.33); Ks--band (2.18 and 0.35), L' (3.8 and 0.62) and M' (4.78 and 0.59). The actual filter transmission curves are available on the ESO instrument website.

Most state-of-the-art ground- and space-based high-contrast observations rely on relative instrument-sky rotations to separate faint point sources from instrument speckles. The ground-based version of this technique is often referred to as angular differential imaging (ADI) -- a technique with a variety of implementations \citep[e.g.][]{marois06a, apai08, kasper07,lafreniere07}. While a powerful method, ADI-type observations necessarily pose an important constraint on time-series observations. Because the observations rely on field rotation to separate real sources from speckles, for each independent photometric measurement a minimum field rotation rate per image (an upper limit on cadence) is required.  We describe this by requiring the arc traced by the planet during the rotation to be larger than at least four times the full width at half maximum of the point spread function (PSF) to ensure that the planet's PSF is well separated from instrument speckles. To calculate the field rotation rate and time required by the ADI we follow the prescription of \cite{mclean97} (Chapter 3, equation 3.14) as follows:

\begin{equation}
\label{equation5}
\omega =  \Omega \frac{\cos(A) \cos(\phi)}{\sin(z)}
\end{equation}
\begin{equation}
\label{equation6}
f_{rot} = \frac{4 \frac{\lambda}{D}}{r_{sep}}
\end{equation}

where $\omega$ is the field rotation rate in radians per second, $\Omega = 7.2925 \times 10^{-5}$ radians/sec is the sidereal rate, $A$ is the target's azimuth, $\phi$ is the latitude of the observatory, $z$ is the target's zenith distance, $f_{rot}$ is the minimum field rotation required for ADI in radians, $\frac{\lambda}{D}$ is the size of the PSF in arcseconds and $r_{sep}$ is the radial distance from the axis of rotation in arcsecond. It is important to note that for a 30--m class telescope the gain in cadence is proportional to the primary mirror's diameter (see Eq.~\ref{equation6}). This is because the diffraction-limited PSF is smaller, which in turns makes the necessary arc traced during the ADI rotation shorter. Space telescopes that allow the rotation of the entire spacecraft, such as the Hubble Space Telescope, may in theory allow for a better temporal sampling, but in practice such rotations are often time--consuming and may limit the cadence. The factor of 4 in the nominator of Equation \ref{equation6} comes from the requirement for non-overlapping PSFs. Throughout this paper we use a target with a declination of $\delta \sim +21^\circ$ (reminiscent of HR8799) observed from the latitude of the Paranal observatory ($\phi = -24^\circ$). Using Equation \ref{equation5} for these set of parameters we obtain a value for $\omega$ near meridian of $\sim 0.02 ^\circ /sec$

Due to the complexity of adaptive optics instruments and the sensitivity of the AO correction to the atmospheric conditions it can be challenging to reach absolute photometry with ~1\% accuracy. While extreme AO systems are expected to reach very high signal to noise ratios on bright giant planets, as discussed in Section \ref{8mExAO}, here we also offer three further techniques that will help reaching high-precision photometry. First, we point out that rotational mapping does not necessarily require absolute photometry. In some cases the host star itself can be used as a photometric reference source. This is not always possible due to the high contrast but other planets in the system should provide ideal comparison points. A good example is the HR8799bcde system, where relative photometry between the three, similarly bright outer planets can provide accurate relative measurements \citep{apai12}. Second, most AO high-contrast imaging pipelines allow the injection of an artificial star in the raw data, which can quantify flux losses during the data reduction. Such artficial star tests can be combined with the relative photometry to further quantify losses and uncertainties. Third, we point out that high-order deformable mirrors can be used to inject an artificial "speckle" into the optical system \citep{marois06b}. Such "speckles" are images of the star and thus can serve as ideal references for relative photometry, even if there are no suitable planets or if the star itself is too bright.Therefore, reaching even sub-percent accuracy in relative photometry with next-generation AO systems seems plausible.

We explore six distinct realizations of the different surface types present on the atmosphere of the giant planet, shown in Table \ref{Table3}: a) cloudy spots on a clear surface, with the same (Model A1 hereafter) or with different temperatures (Model A2 hereafter), representing clouds on a cloud-free surface;  b) clear spots on a cloudy surface, with the same (Model B1 hereafter) or with different temperatures (Model B2 hereafter), representing clear, deeper holes in an otherwise global cloud cover; c) clear surface with cold and hot clear spots (Model C hereafter); and d) cloudy surface with cold and hot cloudy spots (Model D hereafter). Using the B06 cloudy and clear models we simulate spectral modulations due to rotation for effective temperatures between 800K and 1200K and a wavelength range between $1$ and $12$ $\micron$. 

In the next section we present the results of our model using a spectral resolution of 100 and the simple spot distribution in Figure \ref{fig1}.

\section{Results}
\label{Results}

The model described in the previous section allows us to predict rotational variations in the photometry and spectroscopy of giant exoplanets and simulate their observations with existing and next--generation facilities. Here we will evaluate the capabilities of different telescope/instrument classes for characterizing giant exoplanets beyond one--dimensional measurements. We also identify the ideal instrument setup as a function of target temperatures and cloud properties. 

\subsection{Properties of the Simulated Variability}
\label{Section3.1}

An example of normalized model lightcurves is shown in Figure \ref{fig4} where we plot the results for the J, H, Ks and L' filters for Model B2, as described in Section \ref{Section2.3.2}. The maximum normalized amplitude in all three bands occurs when the giant, Clear hot spot (which is 200~K hotter than the 1,000~K effective temperature of the Cloudy surface) rotates into view. As expected from a closer inspection of the green and yellow lines in the top panel of Figure \ref{fig3}, the largest photometric variations, up to 19\% from the mean occur in the J-band. This behavior is consistent with results from \citet{artigau09} and \citet{radigan12} who also reported the largest photometric modulations in the J band for two early T--type brown dwarfs. The H and K-bands have similar behavior in Figure \ref{fig4} with 14\% maximum photometric amplitude for the former and 15\% for the latter while the amplitude of the modulation in L' is not more than 10\%. The minimum in all bands is caused by the two Clear (red spectrum, top panel of Figure \ref{fig3}) cold (T$=$800~K) spots seen on the right panel of Figure \ref{fig1}. This lightcurve minimum is prominent -- up to 13\% lower than the mean J band lightcurve -- given that neither of the cold spots cover more than 4\% of the visible hemisphere (see Section \ref{Section2.1} and Figure \ref{fig2}). Note that the trough is not a symmetric feature because the giant spot rotates into view and dominates the lightcurve, decreasing the effect of the two cold spots. There are many small spots in the model (less than 1\% of the visible hemisphere covered) that contribute small-level photometric variations, whose presence can be deduced from the lightcurve with sufficient sampling rate and accuracy.

Simulated normalized rotational spectral modulations for Model B2 are shown in Figure \ref{fig5} for the JWST/NIRSPEC and JWST/MIRI wavelength range with a spectral resolution of 100. The variations are very strong (up to 35\% minimum to maximum) not only in the J, H and Ks--bands but also in the mid--infrared (MIR) where they can be as high as 20\% at 6 $\micron$. The rotation phases for the rise and fall of the amplitudes in the MIR show distinct variations, in particularly at 9.7 $\micron$ compared to 10.3 $\micron$. The heterogeneous nature of the atmosphere is easily discernible as spots of different temperature and chemical composition rotate in and out of view. Such spectral maps will not only suggest variations in the abundance of gas--phase absorbers, in particular methane or water, but will also further constrain the covering fraction and longitudinal distribution of the spots. 

To further explore the behavior of spectral maps and to study how they respond to changes in the input parameters (surface type, temperature, covering fraction), in Figure \ref{fig6} we present three different realizations of the spectral map in Figure \ref{fig5} for wavelength range between 1 and 2.5 $\micron$. Here we change only the temperature of the spots and their type (Cloudy or Clear) but keep the $T_{eff}$ = 1,000K and total spot-covering fraction $f_{c}$ constant at 10\% and using the atmospheric map in Figure \ref{fig1} throughout. Changing the covering fraction will change the amplitude of the variations, but not the wavelengths/spectra or the periods. Using the nomenclature of Table \ref{Table3}, we show Model A1, Model A2 and Model C on the top, middle and lower panels respectively. The relative variations in the J, H, and Ks-bands are different between the different scenarios. When the giant spot rotates into view, the J-band modulations for Model A1 are negative (as there is less of the brighter, Clear surface; green line in top panel of Figure \ref{fig3}), while those in H- and Ks-bands are positive (as there is more of the brighter, Cloudy surface; yellow line, Figure \ref{fig3}). On the contrary, the J-band variations are positive for both Model A2 (middle panel) and Model C (lower panel) when the giant spot faces the observer. The alternating darker and brighter stripes in the J band in the middle panel of Figure \ref{fig6} are a direct consequence of the differences between the 1,000~K Clear (top panel Figure \ref{fig3}, green) effective surface and the 1,200 K Cloudy (top panel Figure \ref{fig3}, light orange) model spectra -- the former is higher but narrower compared to the latter in this waveband. When the Cloudy Spot rotates into view, it enhances the flux density (white, J-band in the middle panel of Figure \ref{fig6}) in the sides and decreases it in the middle of the filter band, where the contribution of the brighter, Clear surface is lower. Therefore, while photometric variations like the ones in Figure \ref{fig4} can be interpreted as caused by either a hotter/colder region and/or different chemical composition, {\em spectral} modulations, such as the ones seen in the J--band in middle panel of Figure \ref{fig6}, can identify gas-phase compositional variations. The magnitudes of the variations are different between the three models: the variations are strongest in Model C (lower panel, minimum to maximum amplitude of 20\%) and weakest in Model A1 (top panel, minimum to maximum amplitude of 10\%). We note that the magnitudes of the variations for all three models in Figure \ref{fig6} are significantly smaller than those in Figure \ref{fig5} where the minimum to maximum amplitude is up to 35\% (for model B2 from \ref{Section2.3.2}). Additionally, detection of such spectral modulations can also further improve light curve inversion techniques such as the principal component analysis discussed in Section \ref{section4}.

Next we explore the optimal filter for detecting variations for Model B2 as a function of the temperature of the giant spot. In Figure \ref{fig7} we show the results for three different effective temperatures of the planet -- 800~K, 1,000~K and 1,200~K (top, middle and lower panels, respectively) for the J, H and Ks filters. The temperature of the giant spot varies between $-$300K and $+$300K from the effective temperature with $\delta T = 100K$, while the temperatures of the rest of the spots (either cooler or hotter by 200~K compared to the effective temperature) stay the same throughout. Changing the temperature of the giant spot while keeping everything else constant will not only change the amplitude of the variations, but also the shape of the lightcurve with the result that different features will produce the largest amplitudes. The two cold spots in the right panel of Figure \ref{fig1} are very prominent in some cases where the giant spot plays only a secondary role (for example, $T_{eff} = 1200$K and $T_{spot} = 1000, 1100, 1300$K) and are in fact responsible for the largest amplitudes. Regardless of the features responsible, the largest photometric modulations -- the variations producing the strongest observable signal -- occur in the Ks--band for $T_{eff}=800K$ (up to 60\% for the hottest spot, upper panel, Figure \ref{fig7}) and in the J--band for $T_{eff}=1000K$ and $1,200K$ (up to 30\%, middle and lower panels, Figure \ref{fig7}). This result suggests that for sources with physics/chemistry similar to those in Model B2, like SIMP0136 \citep{artigau09} and 2M2139 \citep{radigan12}, the most appropriate filter where the photometric variations are largest and will be most easily detected is indeed J, as these authors noted. For colder sources or sources with a cold spot, however, Ks would be more appropriate. As expected, all three bands "brighten" as the temperature of the giant spot increases from $T_{eff}+100K$ to $T_{eff}+300K$, reminiscent of the well-known J--brightening effect \citep{burgasser00, burgasser02, leggett00} probably caused by the appearance of cloud--free regions at the L/T transition. Here, the brightness increases not due to increase in the size of the clear regions, but due to the increase in their temperature. The three bands brighten at different rates, from the slowest increase in amplitude of 12\% (Ks--band, lower panel) to the fastest of 40\% (Ks--band, upper panel). The fastest brightening for all three filters consistently occurs for the model in the upper panel. We also note the continuous transition in the most appropriate choice of filter where the largest amplitudes occur. As the $T_{eff}$ decreases from 1200K to 800K the best filter changes from J (lower panel) to Ks (upper panel) and as the temperature of the giant spot decreases from $T_{eff}+300K$ to $T_{eff}-300K$, the best filter changes from J to Ks (middle and lower panels). With even larger temperature differences, Ks--band amplitudes would eventually exceed those seen in the J--band. Such large temperature differences between the different surface types are, however, not expected (although see \cite{skemer12}). All the results discussed above are the direct consequence of the interplay between the spectral differences between the Clear and Cloudy model for these temperatures (see Figure \ref{fig3}) and the distribution and sizes of the different surface features. The largest photometric amplitudes occur in different filters for different models. For example, for Model A2 the largest modulation for the same calculation as done for Figure \ref{fig7} consistently occur in the Ks--band. A survey over targets with different $T_{eff}$ can populate Figure \ref{fig7} with more points and in combination with the model presented here can provide information on this level of details.

The following four subsections describe the outcome of simulated photometry for different instruments classes using the theoretical lightcurves from Figure \ref{fig4} and outline the increasingly complex picture of the optical appearance of the giant exoplanet that can be learned. 

\subsection{8--m Class Telescopes with AO}
\label{8mRegAO}

The general prediction of our model is that the largest amplitude changes occur in J-band. However, J-band is not the optimal wavelength for current 8~m-class telescopes with state-of-the-art AO systems:  the planet--to--star flux ratio is $\sim1.2\times10^{-5}$ (see Section \ref{Section2.3.1}). Instead, longer wavelengths provide a better AO correction and more favorable planet-to-star contrasts. We use here the example of VLT/NACO and assume residual stellar PSF in Ks similar to that in L' \citep{kasper07,kasper09,lagrange10,bonnefoy11} to simulate Ks--band photometry where the planet--to--star flux ratio is higher (the planet is brighter): $\sim3.7\times10^{-5}$ (Section \ref{Section2.3.1}). For the Ks--band VLT/NACO PSF FWHM of $\sim$ 0.069 $\arcsec$ we calculate $f_{rot}$ $\sim$ $16^\circ$ from Equation \ref{equation6}. Using this value for $f_{rot}$ and a field rotation rate near meridian $\omega$ $\sim$ $0.02{\frac{\circ}{sec}}$ results in $\sim$14 minutes minimum required time for proper ADI reduction. The signal--to--noise ratio (SNR) for the planet is $\sim$ 15; here we adopted a read noise of 46.20 electrons as given by the VLT/NACO Exposure Time Calculator (ETC) and an observing mode similar to \citep{kasper09} but modified for Ks, namely dithered individual 1 min exposures with detector integration time (DIT) of 0.3454 sec (given by the VLT/NACO manual) and 200 DIT per position. The simulated  Ks--band lightcurve for model B2 with added normally distributed photometric uncertainty of 10\% is shown in the top panel of Figure \ref{fig8}. 

Detection of the photometric modulation is not background--limited but is instead limited by the residual stellar point spread function. The smallest detectable Ks--band variability at 3$\sigma$ level is $\sim$ 20\%, on par with our simulated maximum amplitude (see Figure \ref{fig4}). Therefore, this instrument configuration is capable of confidently detecting the Ks--band photometric variations produced by the giant spot in Figure \ref{fig1} and retrieving the rotation period of the planet. It is, however, inadequate to explore the smaller variations (at the $\sim$1\% level) expected from the smaller clouds.

\subsection{8--m Class Telescopes with next-generation Extreme AO} 
\label{8mExAO}

The next-generation of high-order AO systems (extreme AO, or ExAO) will be capable of delivering very high quality correction also at shorter wavelengths, allowing variability searches to focus on the most favorable wavelengths. To evaluate this configuration we repeat our calculation for the minimum ADI cadence but for J--band PSF -- in this case $f_{rot}$ $\sim$ $9^\circ$ and the minimum required time for proper ADI reduction is $\sim$8 minutes.

Following the science requirements for VLT/SPHERE and Gemini/GPI \citep{mesa11,macintosh08} we use a residual stellar PSF of $10^{-7}$ at a separation of 1$\arcsec$, and a read--noise of 10 electrons to simulate the $\sim$ 100 SNR J--band photometry (with added normally distributed noise of 1\%, see below) for model B2 shown in the middle panel of Figure \ref{fig8}. We note that the high Strehl ratio provided by the Extreme AO (up to 80\%) reduces the speckle noise to such low levels that the limiting factors for the detection of photometric variability will be the photometric precision, read noise and instrument stability. Here we use a 1\% J--band photometric precision following the prescription of \citet{vigan10} for VLT/SPHERE, thus setting the limits on the smallest detectable variability. This instrument configuration will be able to easily measure the rotational periods of planets, will allow preliminary exploration of cloud colors and can even detect the temporal evolution of the cloud cover. Evidence for such changes in cloud cover in brown dwarfs have been reported \citep{artigau09,radigan12}, but their detection requires high signal--to--noise.

For example, the amplitude difference in the simulated (H$-$Ks) color for the giant spot in the right panel of Figure \ref{fig1} is on the order of $0.5\%$ while the (J$-$Ks) amplitude color difference is up to $4\%$. Amplitude difference in the (J$-$Ks) color for the two cold spots in the right panel of Figure \ref{fig1} is $\sim$$2\%$ while that in (H$-$Ks) is $\sim$$0.5\%$. Here we expect a few $\sigma$ detection of (J$-$Ks) modulations caused by the rotation of the giant spot but in general the confidence with which such measurements could be obtained will be ultimately dependent on the photometric precision. A limitation on the capabilities of this class of instruments will be the achievable cadence due to the ADI requirements as discussed in Section \ref{Section2.3.2}, an obstacle that will be overcome by the class of instruments presented in the next section.

\subsection{30--m Class Telescopes with next-generation Extreme AO (30m+ ExAO)}
\label{30mExAO}

This group of instruments represents the suite of planned extremely--large-aperture ground--based telescopes. Following the method described in the previous two sections, we calculate a J--band ADI rotation time requirement of $\sim$ 2 minutes and $\sim$ 10 minutes for a planet at a radial separation from the parent star of $1\arcsec$ and $0.2\arcsec$ respectively; here we present the results for the latter. Using the required characteristics of the GMT/TMT/ELT \citep{gmt06,macintosh06,kasper10a}, we use a residual stellar PSF of $10^{-8}$ at the radial separation of $0.2\arcsec$ to simulate the very high SNR (on the order of several thousands, using the ELT ETC provided by ESO) J--band lightcurve for model B2 in the lower panel of Figure \ref{fig8}. As was the case for the 8m ExAO systems, the residual stellar PSF is no longer the main culprit for systematic errors. Here we adopt a photometric precision of 1\% following the discussion in \cite{dekany04} and \cite{epics10}.

The capabilities of these instruments represent a significant step forward from those of the current 8-- and 10--m class telescopes. The high Strehl ratio combined with the improved cadence provided by the 30--m class ground-based telescopes will allow precise measurements of the basic characteristics of the planet's atmosphere like rotation rate or large--scale inhomogeneity. As for the 8--m class Extreme AO instruments, the 30--m class Extreme AO instruments will also be able to measure cloud colors only down to a level ultimately limited by systematic trends in the photometry, due to the instrumental changes, atmospheric variations, etc. The most significant improvement of the 30--m Extreme AO over the 8--m Extreme AO instruments will be in their ability to do moderate--resolution (R $\sim$ 100 -- 1000) spectroscopy both in the near IR with an Integral Field Spectrograph (IFS) on ELT/EPICS, TMT/NFIRAOS and GMT/NIRExAO Imager \citep{verinaud10,herriot06,johns08} and in the mid--IR using, for example, instruments like GMT/TIGER in the 8 to 20 $\micron$ wavelength coverage \citep{jaffe10}. Further improvements will come in the form of polarimetric measurements and the extended time--domain probed with the faster cadence, allowing detection of smaller-scale variations. Also, the radial separations at which giant exoplanets will be observed with a contrast of up to $10^{-9}$ will be as small as  $0.1\arcsec$, allowing the observations of colder and/or older giant planets in reflected light at orbital distances comparable to that of Jupiter \citep{kasper10a} and of very young planets close to the snow line. Access to such small angular separations will also significantly expand the available distance over which systems containing self-luminous Jovian planets can be directly imaged. This illustrates the unique capabilities of the next--generation extremely--large telescopes as even the future space-based telescopes like the JWST (discussed in the next section) will not have such small angular resolution. Direct imaging of Neptune--size or even Super Earths around the closest stars should be also achievable \citep{kasper10a}. The favorable contrast between such planets and their host star at the mid--IR wavelengths will, albeit through challenging observations, allow low--resolution spectral characterization of their atmospheres by detecting H$_{2}$O, CH$_{4}$, CO,  CO$_{2}$ and NH$_{3}$ features. Such measurements could also discern between a hydrogen--rich and a water--rich atmosphere and test for presence of hazes and non--equilibrium chemistry in the atmospheres of directly--imaged planets (e.g. \citet{miller_ricci11,rogers10, desert11}). Finding such planets in the habitable zone and studying their spectra for biosignatures will also significantly advance the field of astrobiology. 

\subsection{JWST}
\label{JWST}

The James Webb Space Telescope will open a new chapter in the studies of directly-imaged giant exoplanets. With an aperture and Strehl ratio on par with the VLT/SPHERE but free of the atmospheric aberrations, and combined with a dramatically low thermal background, JWST will have a greatly increased sensitivity and cadence. To simulate J--band photometry from NIRCAM we use a residual stellar PSF of $10^{-7}$ at a separation of $1\arcsec$ using the results of \cite{green05}. To reduce the effect of aberrations, we assume a roll--deconvolution \citep{gardner06} mode to be employed by the JWST using a conservative cadence of 15 minutes. Also, following the discussion in \cite{lagage10}, we assume a photometric precision of $\sim10^{-4}$. This is already at least an order of magnitude higher than the one for the ground-based facilities discusses in Sections \ref{8mExAO} and \ref{30mExAO} but is still significantly higher than the achievable contrast, and is again what ultimately limits the level of detectable variations. 

The simulated observation is shown in Figure \ref{fig9}. The significant decrease in the speckle pattern, combined with the very low background noise, the stability of the instrument and the short cadence will allow detailed studies of the rotation periods, cloud distribution, cloud colors and spectra, atmospheric maps and possibly weather patterns of directly imaged giant planets. The unique strength of JWST will be in the complementary observations in the near- and mid-infrared, studying wavelengths that are difficult to explore from the ground, such as the peak in thermal emission of young Jovian planets at 4.5 $\micron$. As discussed by \cite{clampin08}, both the capabilities of NIRCAM in the F460M filter at separations larger than $0.6\arcsec$ and the sensitivity of MIRI will exceed that of even the planned 30-m class ground-based instruments. MIRI will also provide
a platform uniquely suitable for observing planets at wavelengths longer than 5~$\mu$m.

The fact that JWST will not suffer from telluric absorption will also allow more accurate observations of several key gas-phase molecular absorbers. As seen in Figures \ref{fig5} and \ref{fig6} prominent absorbers such as H$_{2}$O, CH$_{4}$, CO and NH$_{3}$ with very strong spectral features can be studied by NIRSPEC and MIRI in detail, allowing their spatial and temporal distribution among other things to be deduced. As discussed by \cite{marley10} and \cite{bailer08} detection of variability in very temperature--dependent spectral features such as FeH, NaI and KI and/or correlated modulations between these features will further support the assumption of a patchy atmosphere. Detection of strong water absorption, for example, will indicate a hydrogen--rich envelope. The JWST will also be uniquely suited to study the mid--infrared regime, where the spectroscopic variations can be as high as 20\% for the model seen in Figure \ref{fig5}. 

\section{Lightcurve Inversion} \label{Map}
\label{section4}

\begin{figure}
\epsscale{1.0}
\plotone{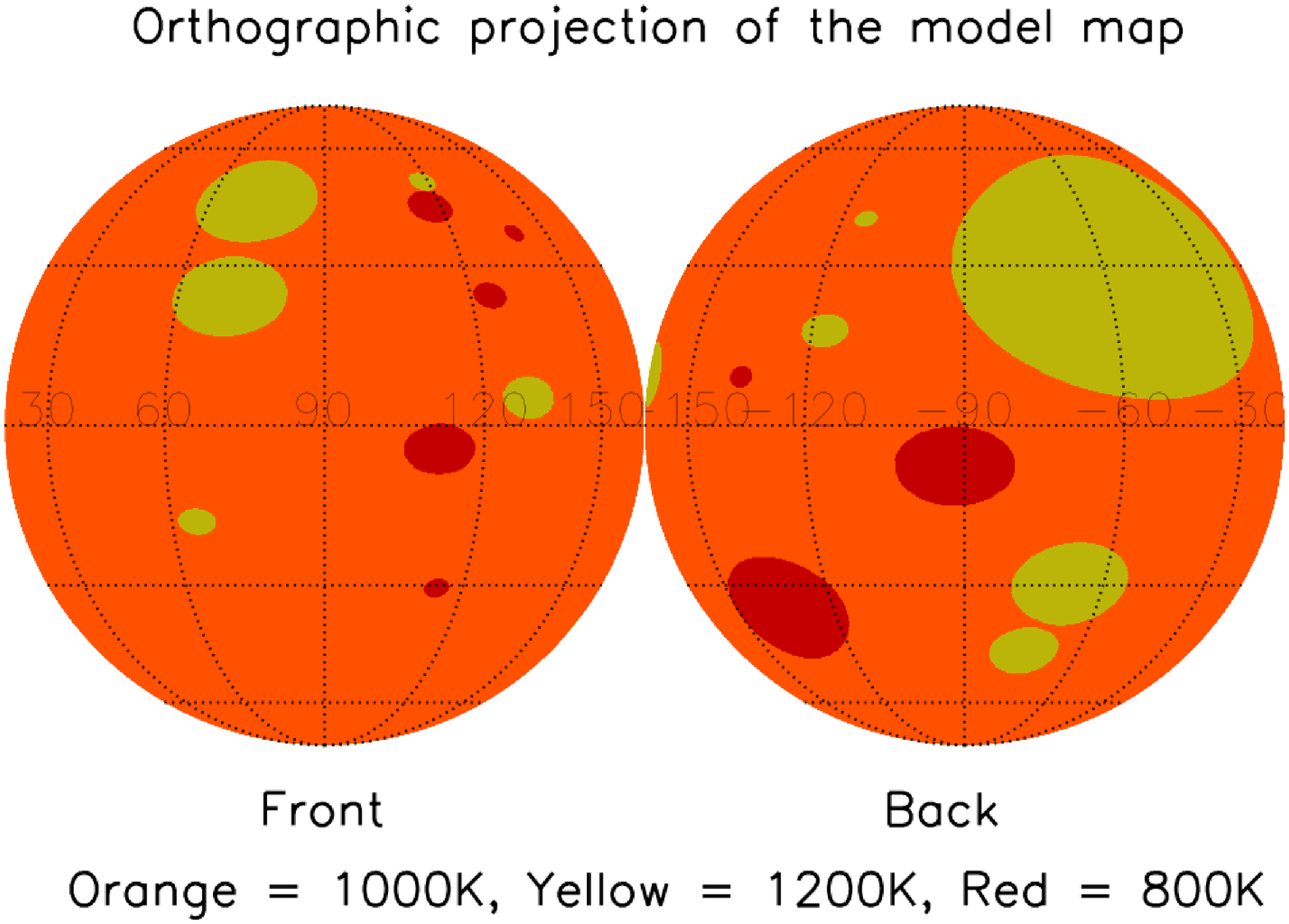}
\caption{Projected model surface map of a giant planet with randomly distributed spots according to a power law with an index of --2.0 (such that the largest five clouds are responsible for 85\% of the signal). Different colors correspond to different surface types temperatures, as indicated in the figure. This model has a 10\% total spot--covering factor (the giant spot covers 5\% of the total surface area), 3 different temperatures, and inclination of 0$^\circ$. The left disk represents the front side of the planet facing the observer, the right disk -- the back.\label{fig1}}
\end{figure}

\begin{figure}
\epsscale{1.1}
\plotone{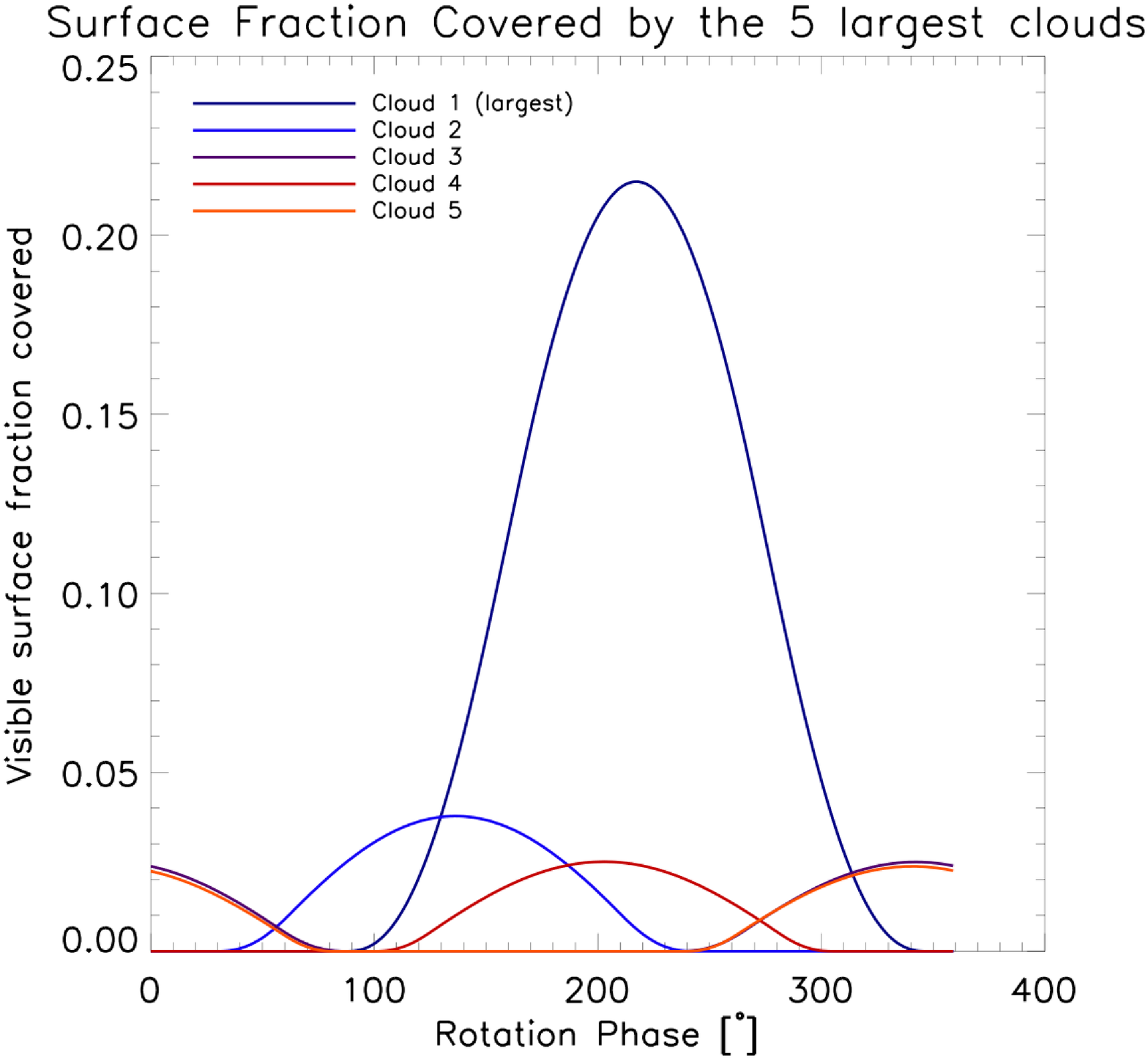}
\caption{The visible surface fraction covered by the 5 largest spots as a function of rotational phase for the map shown in Figure \ref{fig1}.\label{fig2}}
\end{figure}

\begin{figure}
\epsscale{1.0}
\plotone{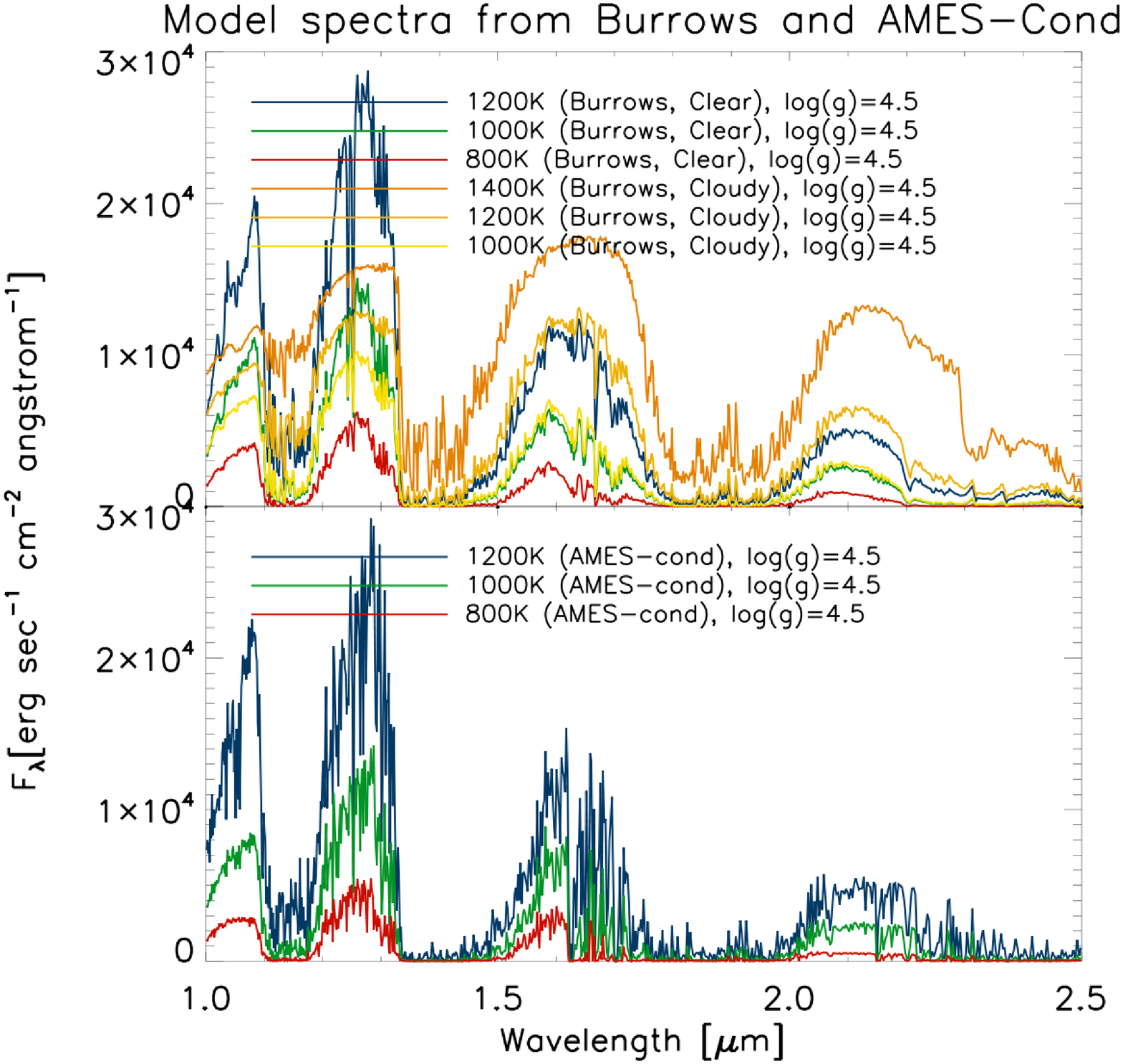}
\caption{Sample spectra for temperatures of 800K, 1000K and 1200K and log(g)=4.5 from the Clear Burrows (B06, top) and AMES-Cond (A01, lower) models.\label{fig3}}
\end{figure}

\begin{figure}
\epsscale{1.0}
\plotone{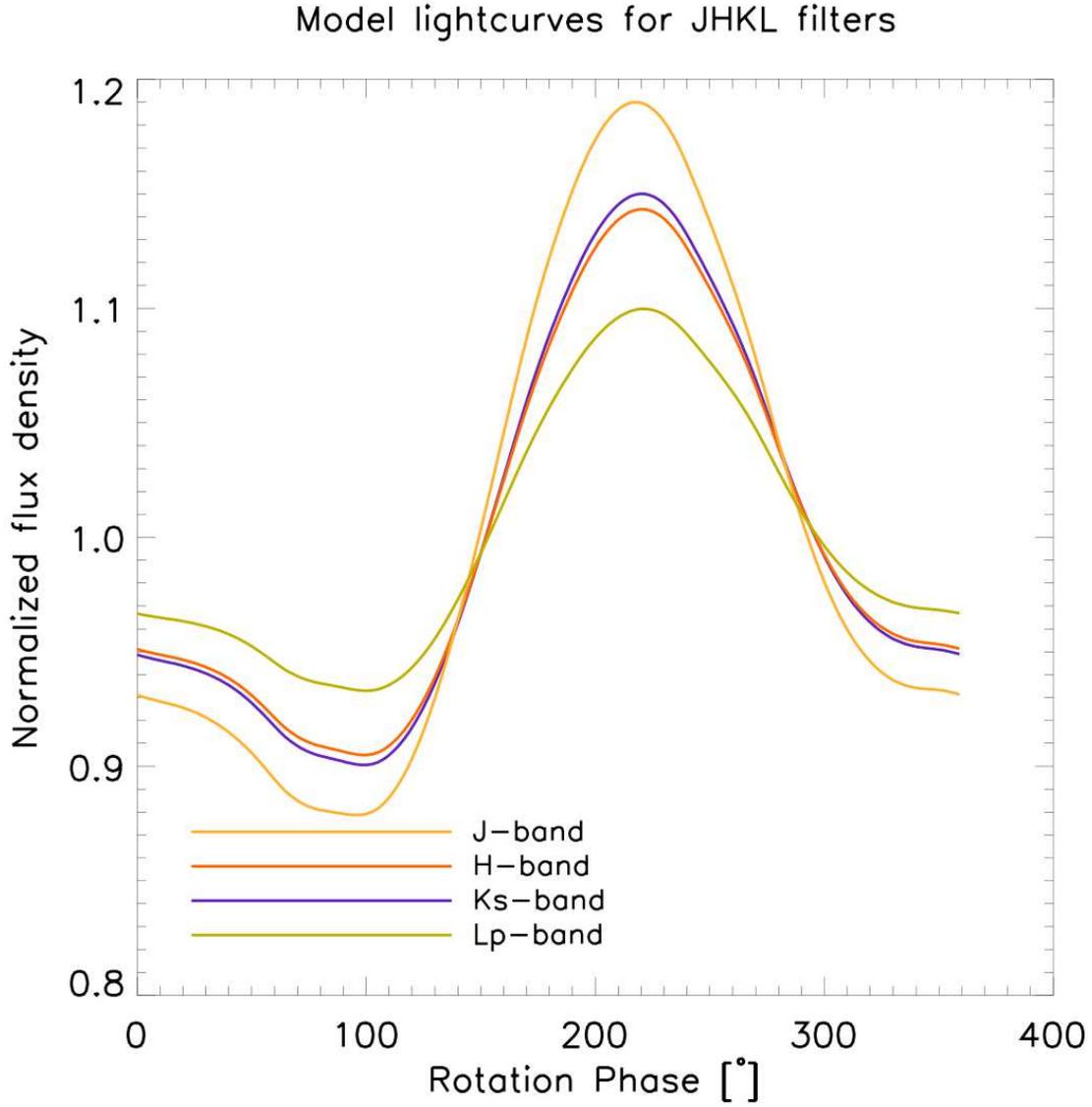}
\caption{Normalized model lightcurves as a function of rotation phase for the map in Figure \ref{fig1}, using Model B2 (clear spots on a cloudy surface, with different temperatures) and assuming $T_{eff} = 1000K$ and the Giant Spot at 1200K. As discussed in the text, a suite of next--generation instruments, dedicated to direct imaging of giant exoplanets, can detect such modulations and measure the rotation periods and/or cloud--coverage of the planet. \label{fig4}}
\end{figure}

\begin{figure}
\epsscale{0.85}
\plotone{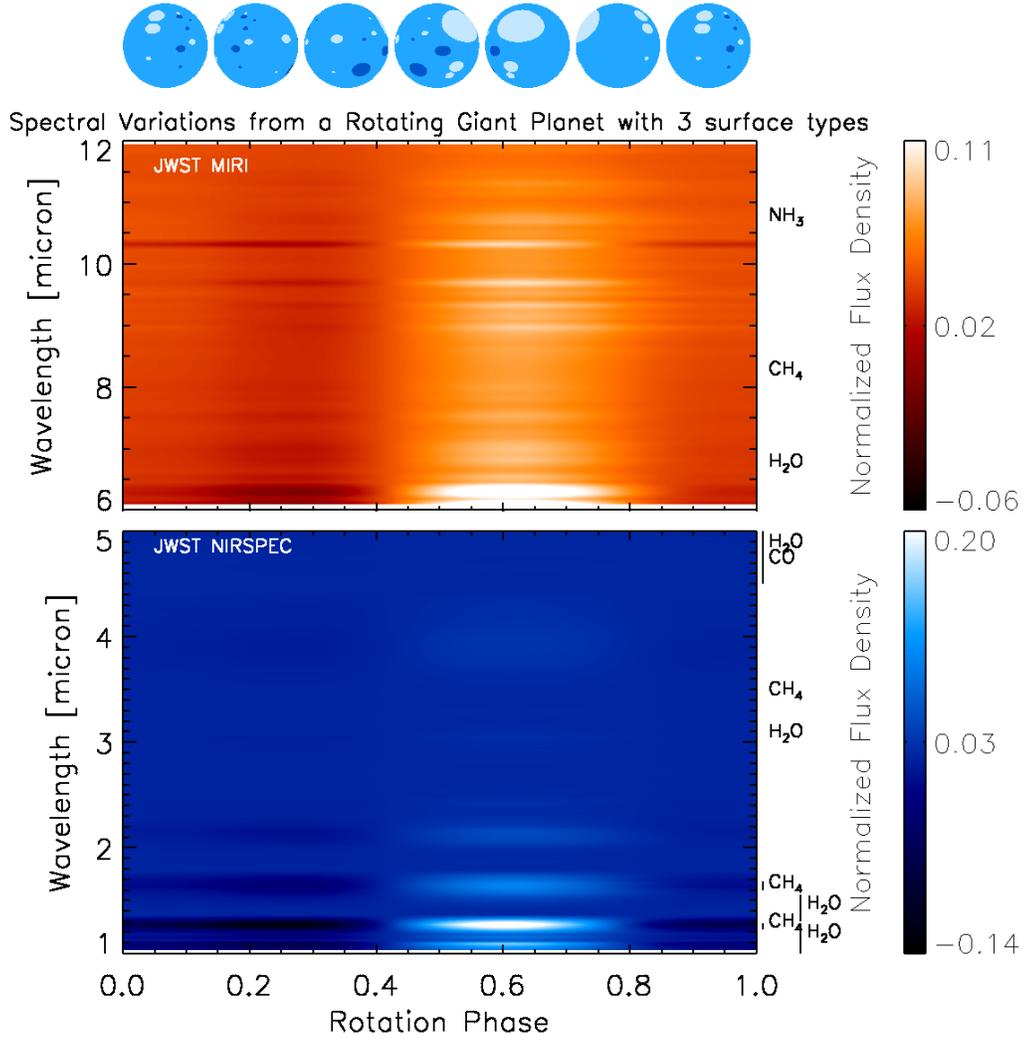}
\caption{Predicted spectral modulations (normalized) for JWST/NIRSPEC and JWST/MIRI (with a resolution of 100) as a function of the rotation phase for the map in Figure \ref{fig1}, using a model with clear spots on a cloudy surface (B2 from Table \ref{Table3}) for a planet with $T_{eff} = $1,000~K and a giant hot spot of 1,200~K. The top panel shows the visible hemisphere as a function of orbital phase (shown at the bottom). Absorption bands for the respective molecules are marked on the right. \label{fig5}}
\end{figure}

\begin{figure}
\epsscale{1.0}
\plotone{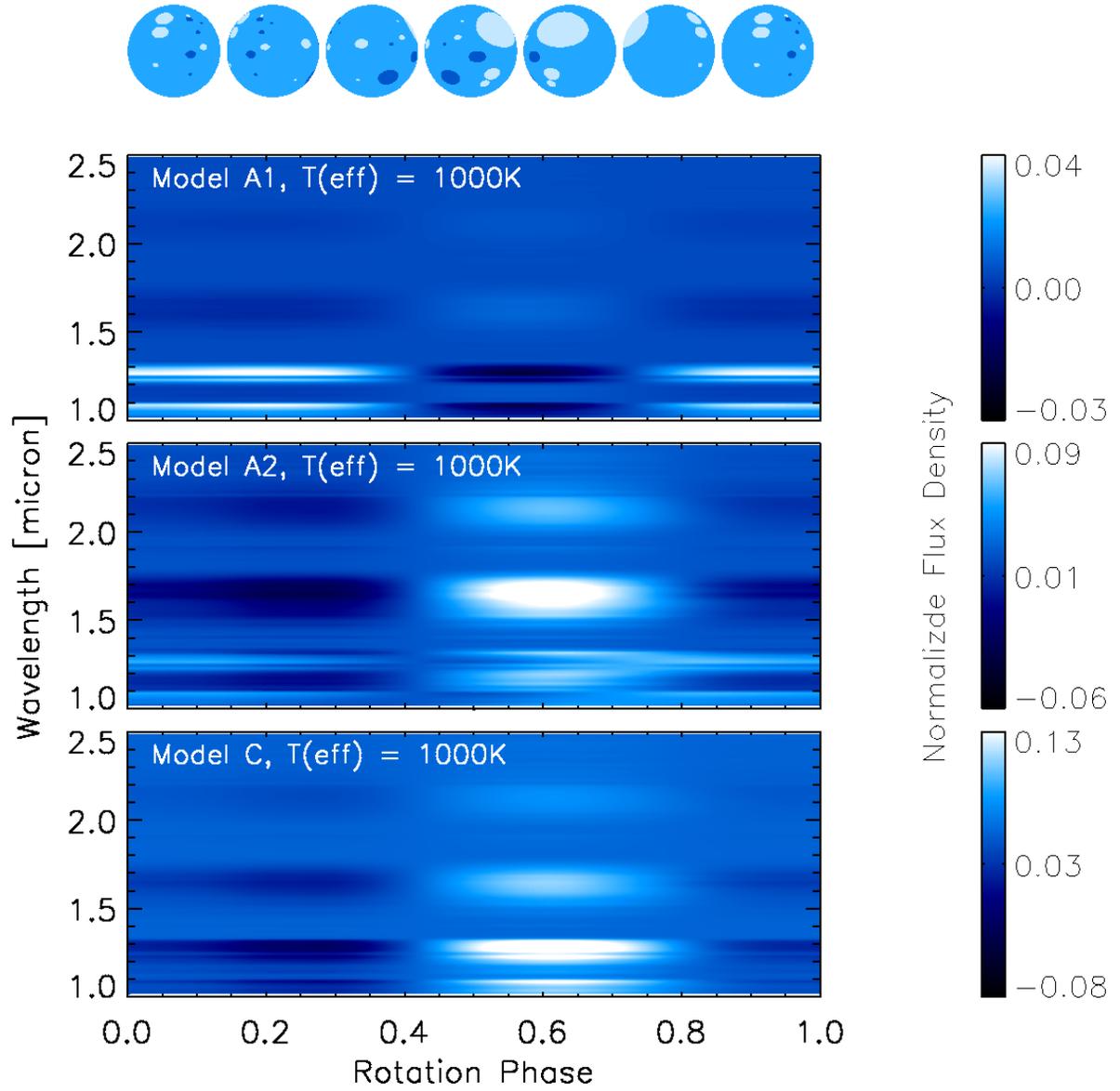}
\caption{Same as Figure \ref{fig5} but for different properties of the spots (temperature and surface type). Top panel is for Model A1, middle panel is for Model A2 and lower panel is for Model C. All models are for a $T_{eff} = 1000K$.\label{fig6}}
\end{figure}

\begin{figure}
\epsscale{1.0}
\plotone{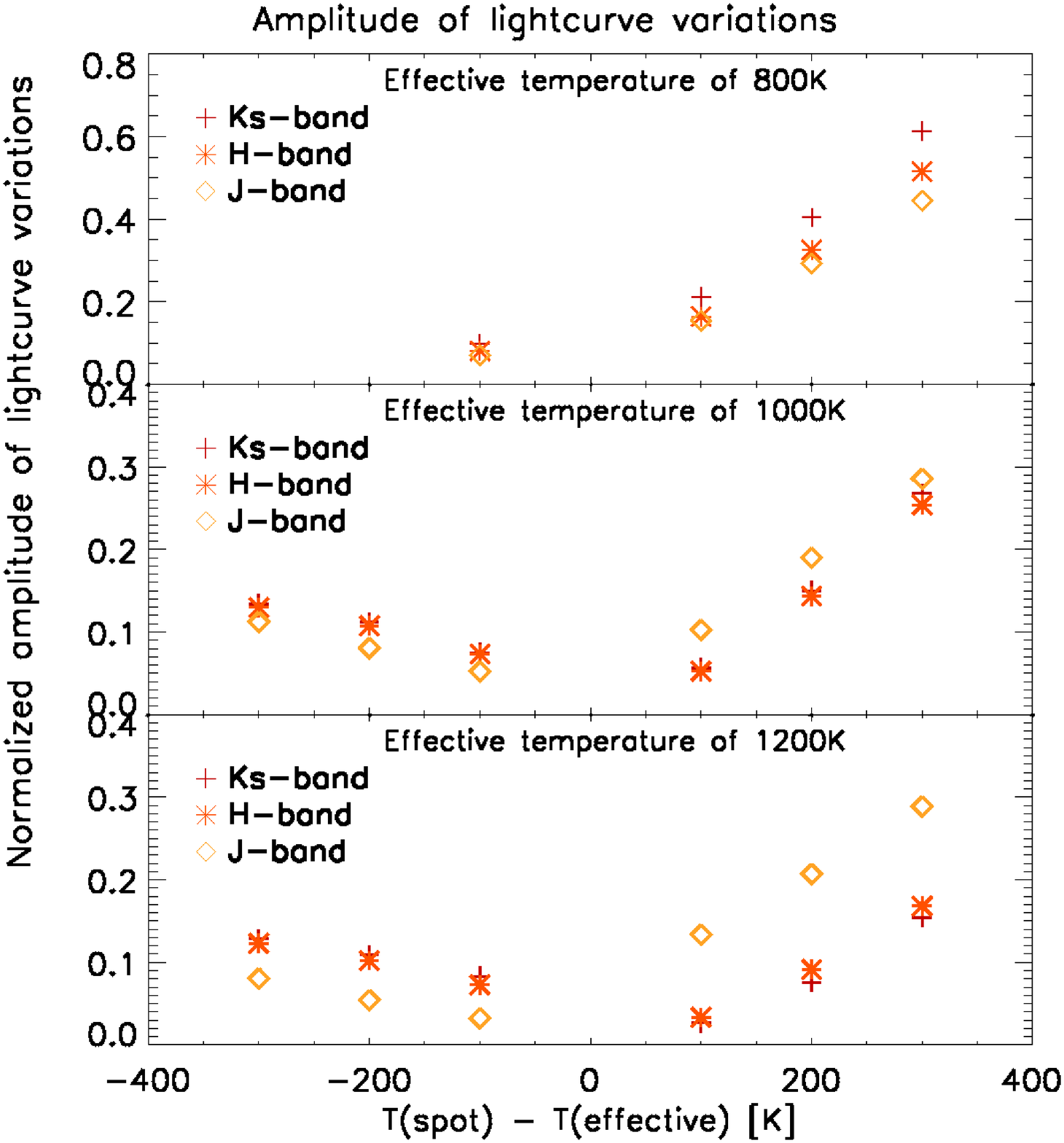}
\caption{Normalized amplitudes of lightcurve variations as a function of the temperature of the giant spot for the map in Figure \ref{fig1}. The top panel is for $T_{eff} = 800K$, the middle panel is for $T_{eff} = 1000K$ and the lower panel -- for $T_{eff} = 1200K$. All three panels are for Model B2 (Table \ref{Table3}). The largest variations occur in different filters depending on the temperature of the giant spot, suggesting that the observations should be carefully tailored to the specific target.\label{fig7}}
\end{figure}

\begin{figure}
\epsscale{1.0}
\plotone{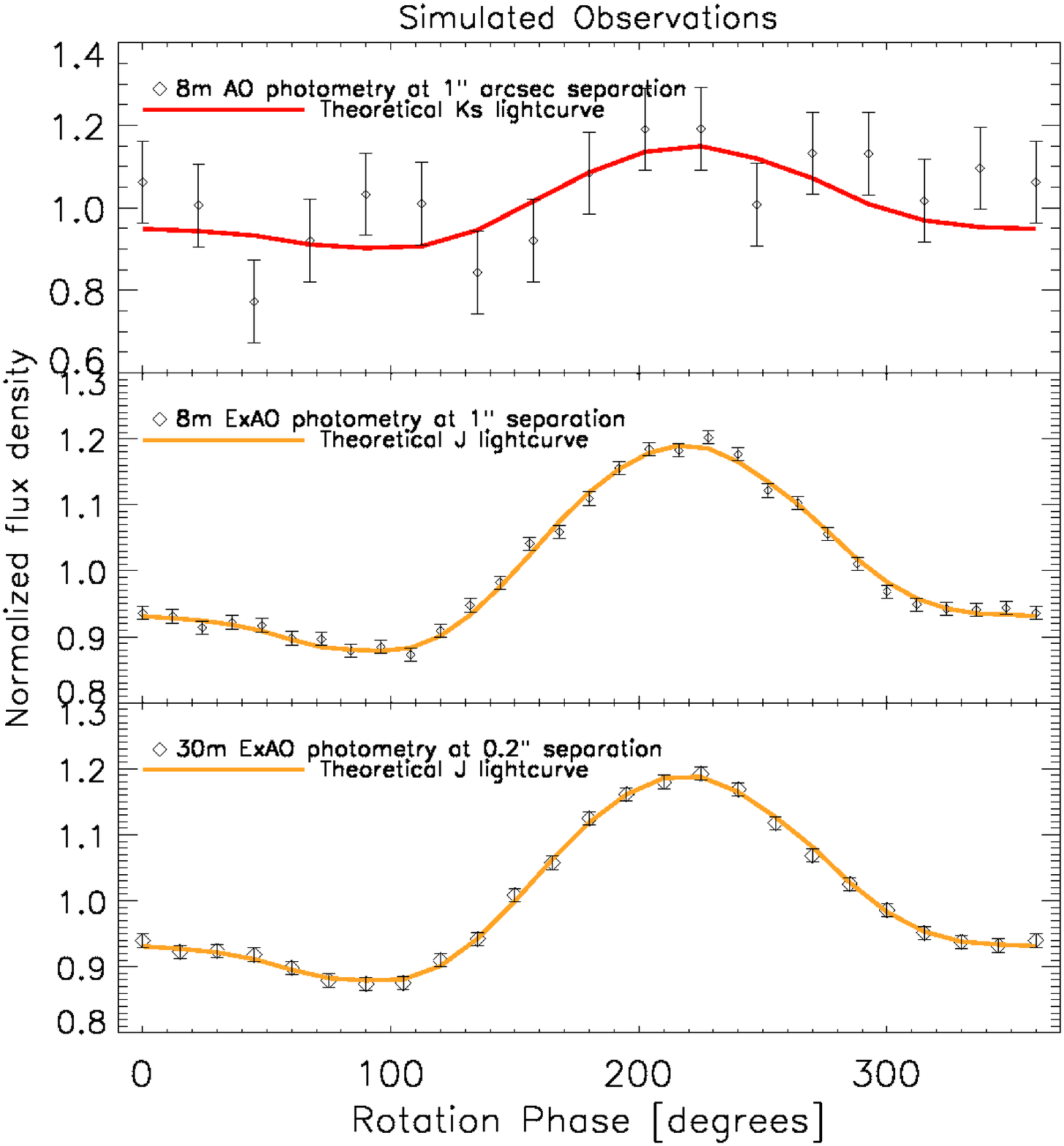}
\caption{Simulated observations of a $T_{eff} = 1000K$ planet with a giant hot spot of 1200K (Model B2) for an 8--m class telescope with AO (upper panel) and with Extreme AO (middle panel), and for a 30--m class telescope with Extreme AO (lower panel). The red and yellow line are the theoretical lightcurves from Figure \ref{fig4}, the black diamonds are simulated photometry; the error bars represent a photometric precision of 10\% in the upper panel and 1\% in the middle and lower panels. We assume a rotation period of 4 hours. The temporal sampling is 15 min cadence for the 8-m AO, 8 min for the 8-m ExAO and 10 min cadence for the 30-m aperture.\label{fig8}}
\end{figure}

\begin{figure}
\epsscale{1.0}
\plotone{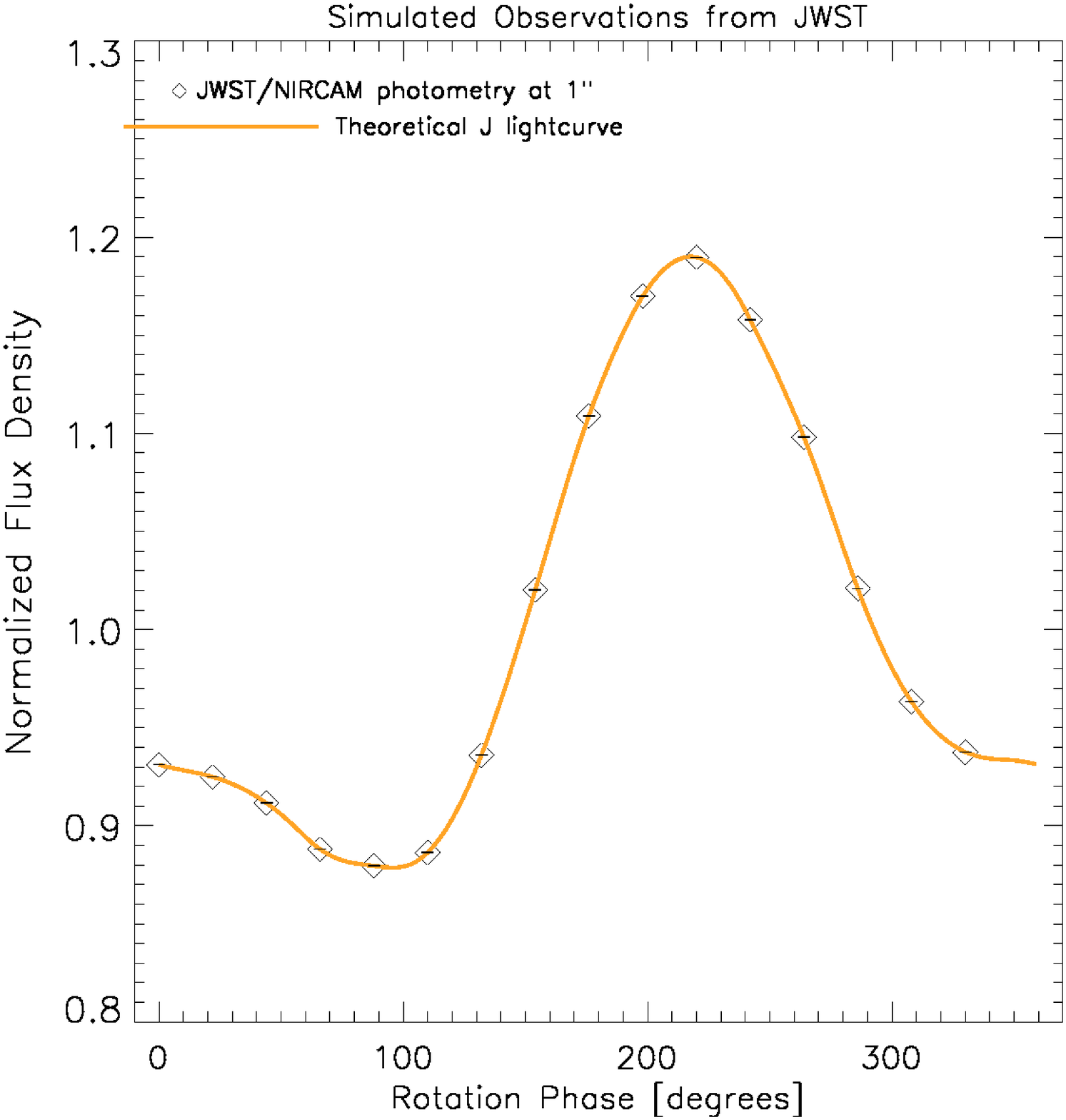}
\caption{Same as the previous figure, but for simulated observations from JWST/NIRCAM F115W with a cadence of 15 min and photometric precision of $\sim10^{-4}$.\label{fig9}}
\end{figure}

To explore the feasibility of recovering the map from Figure \ref{fig1}, we use a principal component analysis (PCA) and sinusoidal variations in surface brightness as a function of only the longitude as studied by Cowan et al. (2009) in combination with the formalism for light curve inversion (LCI) developed in \cite{russell1906}. The purpose of this section is to illustrate the potential of LCI.

\begin{figure}
\epsscale{1.0}
\plotone{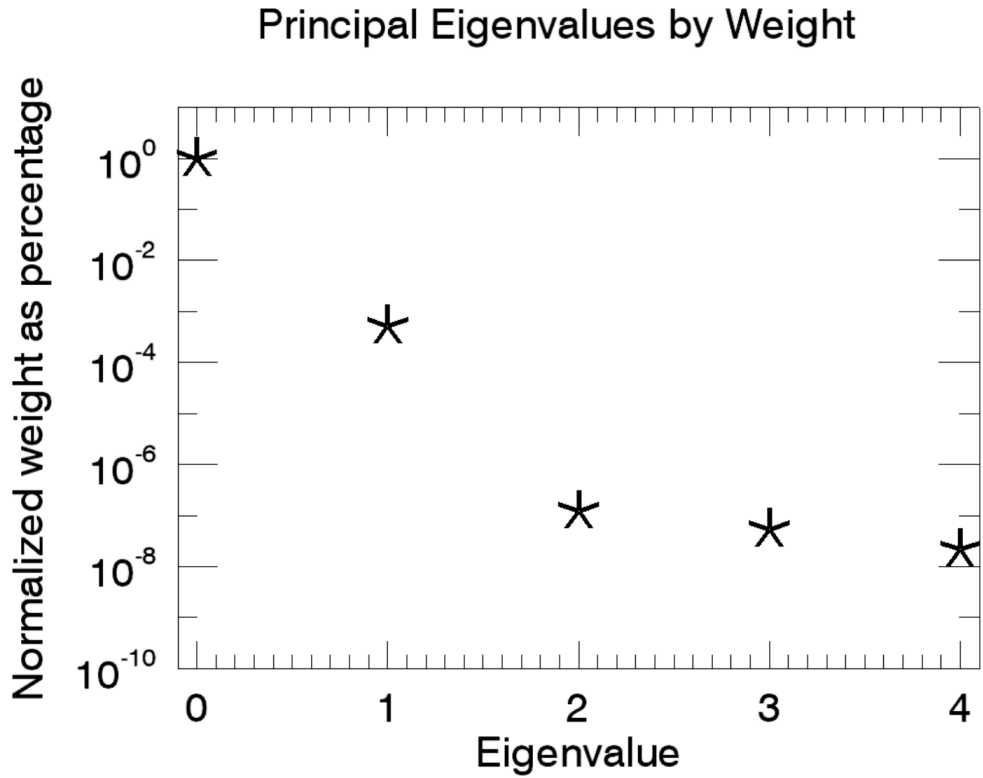}
\caption{Principal eigenvalues by weight: 0.998, 2.2$\times$$10^{-3}$, 2.2$\times$$10^{-7}$, 6.3$\times$$10^{-8}$ and 3.1$\times$$10^{-8}$. The two largest values correspond to the two eigencolors representing the two different spot temperatures we assumed in our model. The results shows that the simulated photometric variations are indeed caused by only two colors (different from the mean color, which is a manifest of the effective temperature of the planet), or precisely the number of "extra colors" in our input map. The other three eigenvalues are zero to the precision we used. We used our own PCA routine, but the built--in IDL PCA routine also gives zeros for these three.\label{fig10}}
\end{figure}

\begin{figure}
\epsscale{0.65}
\plotone{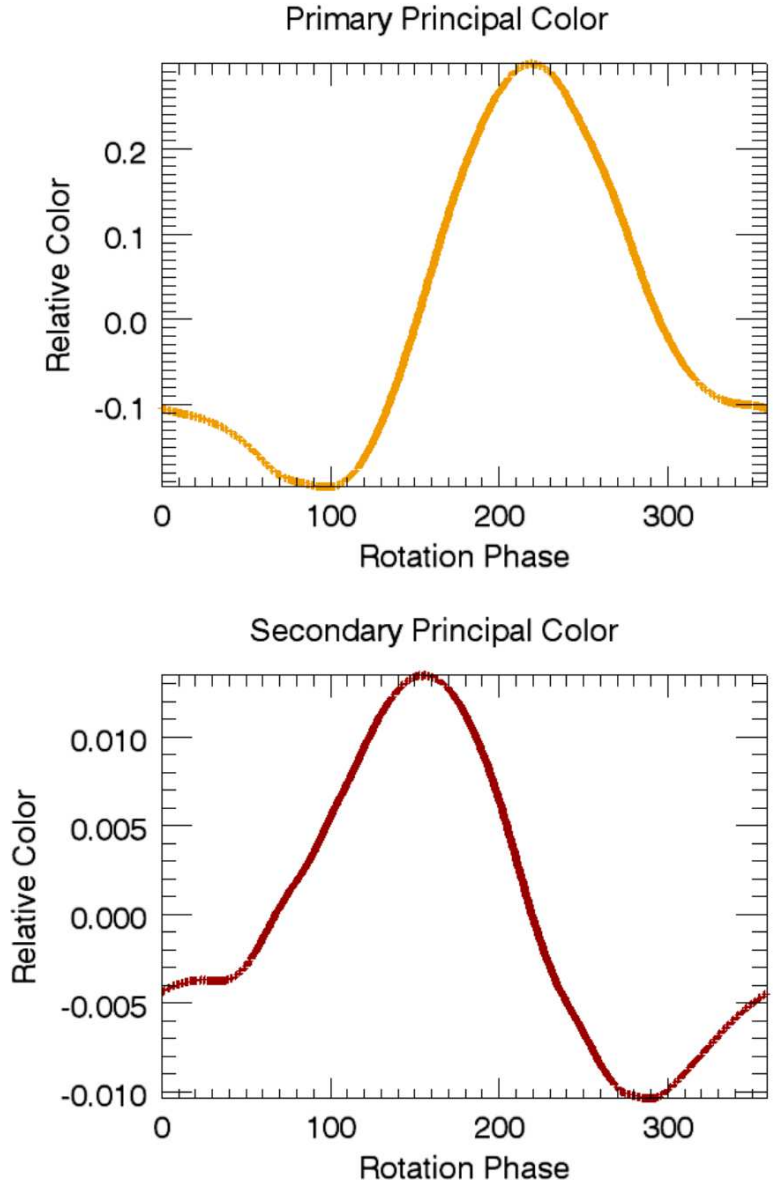}
\caption{Phase variations of the primary (top panel) and secondary (lower panel) eigencolors as inverted by PCA from the simulated photometry from Figure \ref{fig4}. Rotation phase of zero is defined as the left panel on \ref{fig1}. A maximum in the relative color outlines a rotation phase at which the most amount (largest area covered) of that respective eigencolor is present on the side of the planet facing the observer and a minimum -- the least. The primary eigencolor practically mimics the input (orange) J--band lightcurve from Figure \ref{fig4}, while the secondary has a completely different behavior, hard to notice by eye from the lightcurve. \label{fig11}}
\end{figure}

As an example we utilize a covariance-matrix PCA, using the five J, H, Ks, L' and M' filters as the input vectors defining a 5--dimensional parameter space to be converted to a minimum set of eigenvectors sufficient to produce the simulated variations. 
The resulting normalized eigenvalues are shown in Figure \ref{fig10}. Following the prescription for the method, we notice that the largest gap occurs between the second and the third eigenvalues, suggesting that only two principal colors above the mean background color are required to reproduce the observed variations, consistent with the three colors we used for our map in Figure \ref{fig1}. The contribution from the two primary colors as a function of the rotation phase is shown in Figure \ref{fig11} where we use two panels to show the actual magnitude of the variations (rotation phase of zero is defined as the left panel on \ref{fig1}). A maximum value corresponds to a rotation phase at which the largest area covered of that respective eigencolor is present on the side facing us and a minimum -- the least. The primary eigencolor basically traces the input (orange) J--band lightcurve from Figure \ref{fig4}. On the contrary, the behavior of the secondary is not immediately obvious from the input lightcurve. As pointed out in \cite{cowan09}, the eigencolors do not represent an actual color but a deviation from the mean "color", which for our map in Figure \ref{fig1} is the 1,000~K (orange) background temperature.

\begin{figure}
\epsscale{0.95}
\plotone{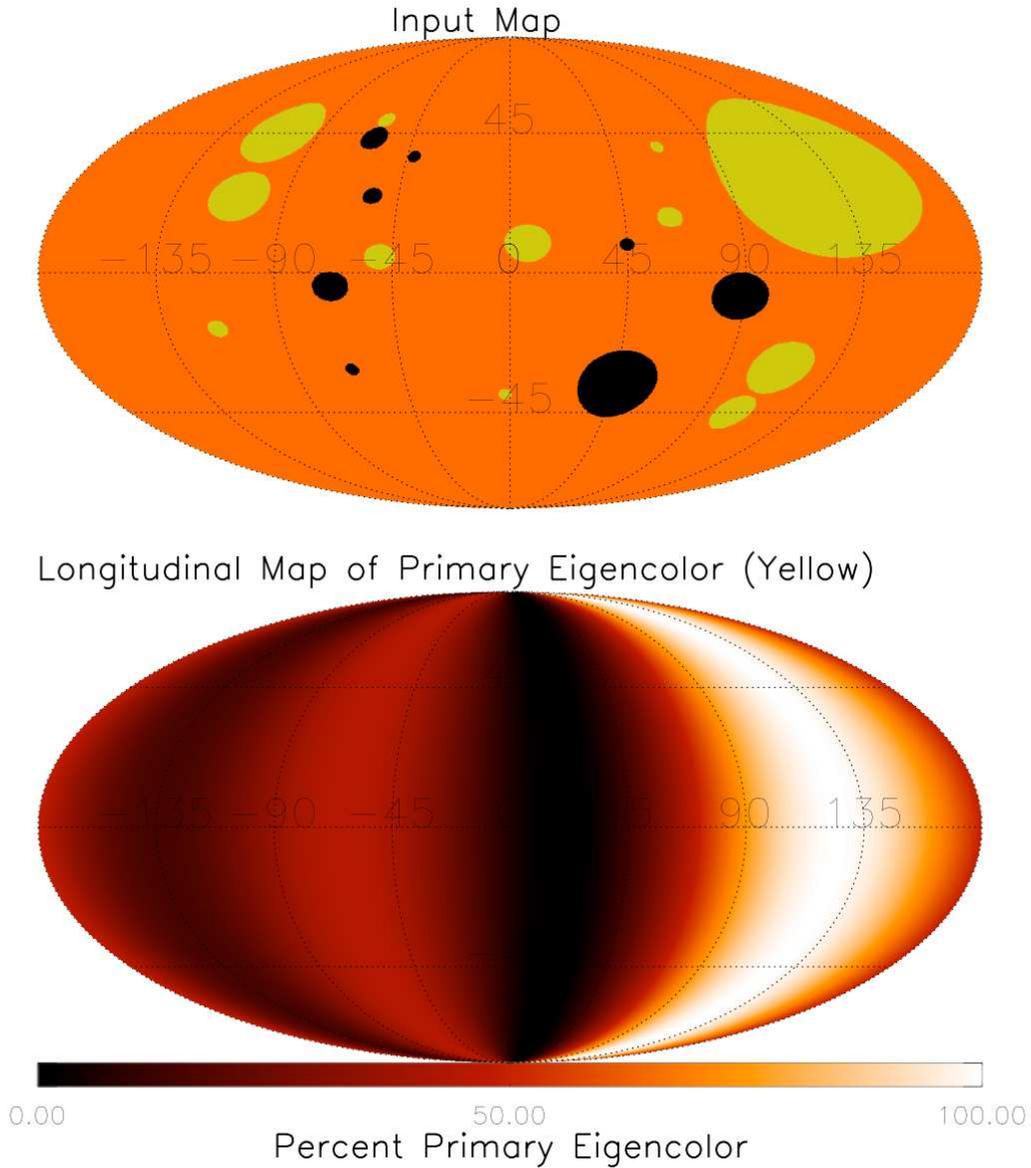}
\caption{Mollweide projection of the input map (upper panel: $orange=1000K$, $yellow = 1200K$, $black = 800K$) and of the PCA--inverted, longitudinal distribution of the primary eigencolor (lower panel: normalized percentage contribution). The inversion method used successfully recovers the longitudinal position (there is no latitudinal resolution) of the giant hot spot at $+$135$^\circ$ and, to a somewhat lesser degree, that of the two smaller spots at $\sim$ $-$110$^\circ$. The secondary eigencolor is masked out (as black) in the upper panel as it is orthogonal to the primary and does not contribute to the variations of the primary.\label{fig12}}
\end{figure}

\begin{figure}
\epsscale{0.95}
\plotone{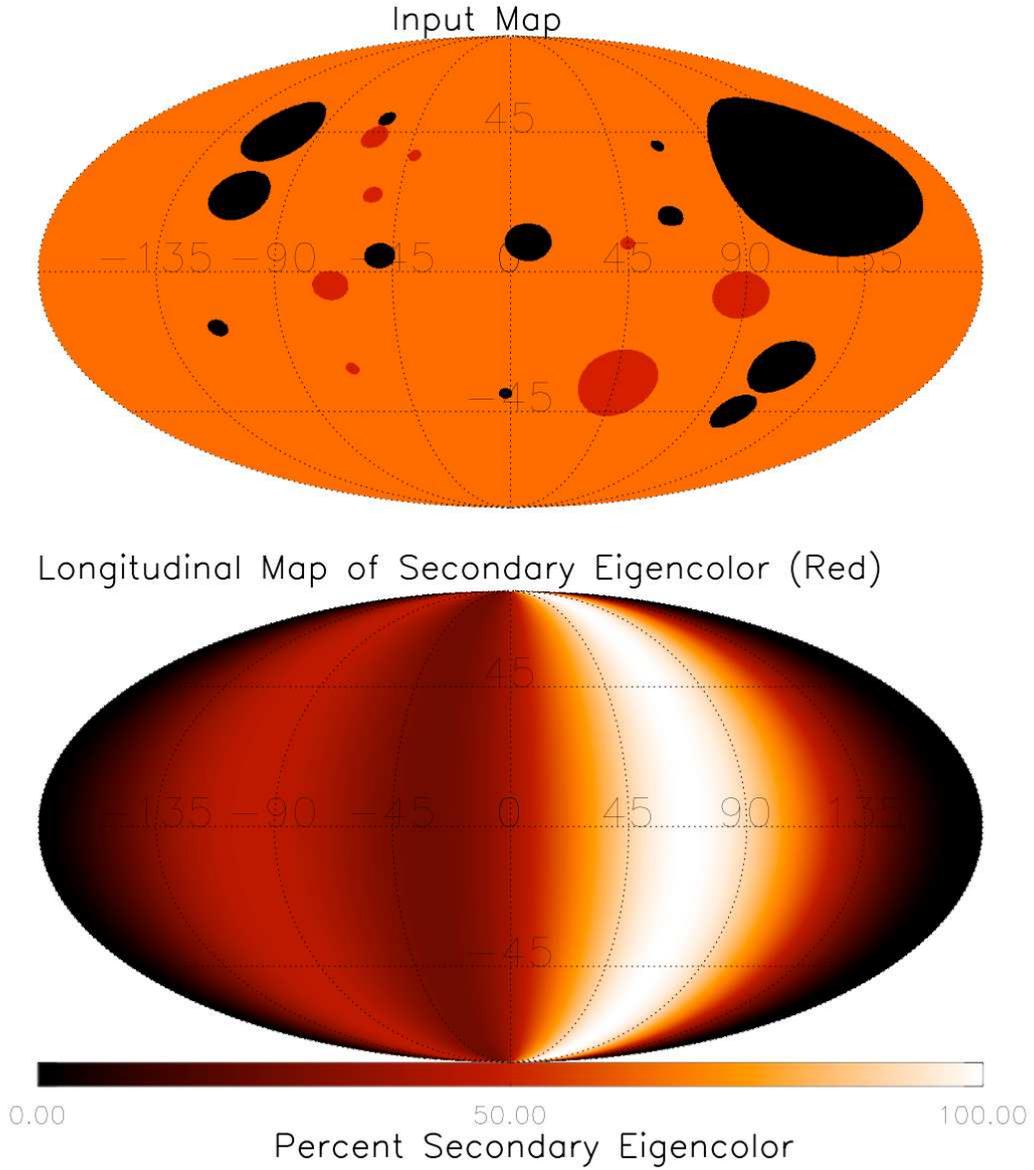}
\caption{The same as the previous figure, but for the secondary eigencolor (upper panel: $orange=1000K$, $black = 1200K$, $red = 800K$). As in the case for the primary color, the inversion successfully recovers the longitudinal position of the two groups of cold spots, at $\sim$ $-$60$^\circ$ and $\sim$ $+$60$^\circ$ respectively. The primary eigencolor is masked out (as black) in the upper panel.\label{fig13}}
\end{figure}

To recover the surface map from Figure \ref{fig1}, we follow the prescription of \cite{russell1906}. We use the two different eigencolors as separate light curves and expand them in spherical harmonics. We then invert them for a longitudinal, sinusoidal brightness distribution. The inversion is done in one dimension only -- there is no latitudinal resolution. For simplicity, we assume that the planet's rotation axis is perpendicular to the line of sight. The resulting longitudinal maps in Mollweide projection for the primary and secondary eigencolors are shown in the lower panels Figures \ref{fig12} and \ref{fig13}, respectively. The "other" eigencolor on each map has been "masked out" because the two eigencolors are orthogonal. In the upper panels of Figures \ref{fig12} and \ref{fig13} we compare the PCA--inverted longitudinal maps to the input map (in the same projection) from Figure \ref{fig1}. The two figures show that this simple method successfully recovers the longitudinal position of the major features, namely the giant hot spot at a longitude of $+$135$^\circ$ (Figure \ref{fig12}) and the two groups of smaller, cooler spots at longitudes of $-$60$^\circ$ and $+$60$^\circ$ respectively (Figure \ref{fig13}). 

As \cite{russell1906} pointed out expanding the lightcurve in spherical harmonics does not completely determine the harmonics expansion of the surface brightness, as all but the first odd harmonics are absent in the lightcurve. Nevertheless, this simple model not only recovers the contribution to the lightcurve from the primary, dominant eigencolor (maximum variation of up to 0.3, upper panel, Figure \ref{fig11}) but also identifies the effect of the much weaker components (maximum variation of 0.013, lower panel, Figure \ref{fig11}) from the secondary eigencolor, allowing us to infer the presence of distinct surface features on the planet.

A more comprehensive way to obtain the surface map of the plane would be to use the PCA on the spectroscopic instead of the photometric variations, using the resolution of the instrument to increase the dimensionality of the data -- a single, low-resolution spectrum of R$\sim50$ will increase the number of orthogonal data sets by an order of magnitude over the five filters discussed above. 

The method described thus far is well suited to study the color asymmetry caused by the spots in the form of spatial distribution over the surface of the planet, after simplifying assumptions like the longitudinal sinusoidal intensity map or an N-slice map (see \citealt{cowan09}). 

\section{Discussion}
\label{Discussion}

\subsection{Patchy Dusty Atmospheres in Brown Dwarfs and Exoplanets}
\label{State_of_the_art}

Early attempts to understand the nature of giant planets were based on the use of atmospheric models of L and T brown dwarfs as templates to assess the detection sensitivity for direct imaging surveys. While at similar temperatures to the regime of the L/T transition objects (between $\sim$ 900K and 1200K where dust clouds settle below the photosphere), it has been suggested that giant planets may, in fact, remain cloudy where the T dwarfs become cloud-free \citep{saumon08,currie11,barman11a,madhusudhan11, skemer12}, reminiscent of the results of \citet{stephens09} on low-gravity BDs. This does not necessary invoke a new kind of atmospheres for the directly--imaged giant planets. Most likely it is an age effect of atmospheres rather more dusty for their spectral type. 

The observed abundance of methane compared to CO in the four planets of the HR8799 system has been reported to be lower compared to chemical equilibrium models predictions, suggesting efficient convective mixing on time-scales shorter than the CO to CH$_{4}$ equilibrium rate \citep{hinz10,barman11a, skemer12}. This assumption is further supported by the 3.88 to 4.10 $\micron$ spectroscopy of HR8799c being inconsistent with both the DUSTY \citep{chabrier00} and the COND \citep{baraffe03} models, indicating non-equilibrium chemistry at work. It appears that giant planets such as HR8799bcde and 2M1207b (a rather more pronounced case) may indeed have the temperature of T--dwarfs but the dusty appearance of L-dwarfs \citep{chauvin04,skemer11,barman11b}. In particular, the luminosity of HR8799b, for example, implies an effective temperature of 850K while its colors suggest a much hotter atmosphere of 1300K \citep{skemer12}. Models with thick clouds (with the cloud base significantly deeper compared to that of L/T dwarfs) and low surface gravity are invoked by \citet{currie11} to match the planets in the HR8799 system. Their best-fit model for HR8799b yields T$_{eff}$ = 800--1000K and log(g) = 4.5 and T$_{eff}$ = 1000--1200K and log(g) = 4--4.5 for HR8799cd.  The authors put less importance on the effects of non-equilibrium chemistry in reproducing the 1 to 5 $\micron$ SEDs but suggest that the planet atmospheres may indeed be out of equilibrium. Interestingly, they also argue that "patchy" cloudy models may provide an even better fit to the data. On the contrary, to explain the observed low effective temperature (less than 1000K, typical of cloud-free T-dwarfs), red colors and smooth spectrum of HR8799b, the results of \citet{barman11a} support the presence of thick photospheric clouds and enhanced metallicity in the presence of non-equilibrium chemistry. The authors note that the spectrum of the planet is markedly different from that of typical field brown dwarfs, namely in the weaker methane and CO and stronger water absorption (suggesting a hydrogen-rich atmosphere). In contrast to the results of \citet{currie11} who stress the importance of cloud thickness over non-equilibrium chemistry, \citet{barman11a} assign equal importance to both. However, their best-fit model for solar abundance, with T$_{eff}$ = 1100K and log(g) = 3.5, indicates a radius that is smaller than theoretically expected, suggesting that the metallicity of the planet must be higher. A chemical enrichment by a factor of ten produces a significantly better fit for their data to a model with T$_{eff}$ = 869K and log(g) = 4.3, consistent with the planet's mass and age. Similar scenario can be drawn for the case of 2M1207b, where the apparent low luminosity and red colors are in stark contrast with the spectrum-derived T$_{eff}$$\sim$1600K. Using an atmospheric model with thick clouds, non-equilibrium chemistry, a mixture of grains and low surface gravity, \citet{barman11b} propose a best-fit solution with T$_{eff}$ = 1000K and log(g) = 4., consistent with cooling track predictions and disfavoring exotic scenarios such as edge-on disks or planetary collisions. For both HR8799b and 2M1207, \citet{barman11a,barman11b} point out that the thick clouds (reminiscent of L--dwarfs) extending across the photosphere, effectively accounting for the observed red colors, and the non-equilibrium chemistry consistent with the observed CO/CH$_{4}$ ratio are equally important but only in the presence of low surface gravity. 

Further observations of the HR8799 planets \citep{skemer12} also indicate that all four planets are brighter than expected at 3.3 $\micron$ compared to equilibrium models that postulate significant methane absorption and dimming at this wavelength. The authors report that the observations are inconsistent with models with decreased CH$_{4}$, thick clouds and non-equilibrium chemistry. Their 3.3 $\micron$ photometry of HR8799b is not consistent with the best-fit model of \citet{barman11a} and is twice the value obtained by \citet{currie11}. Using the models of \citet{madhusudhan11} and adopting a "patchy" cloud cover, \citet{skemer12} find a best-fit model (to all their photometry data except M--band) with a two-component atmosphere of 93\% T$_{eff}$ = 700 K \citep{madhusudhan11}('A'--type clouds) and 7\% T$_{eff}$ = 1400 K ('AE'--type clouds) for HR8799b. Similar hybrid model atmospheres provide better fit to HR8799 c, d, and e compared to thick clouds/non-equilibrium chemistry models that again fail to reproduce the 3.3 $\micron$ colors. The four HR8799 planets have different effective temperatures but, intriguingly, similar colors which led \citet{skemer12} to suggest that their atmospheres should have similar properties and may indeed be an evidence for their patchy appearance. 

Low-level photometric variability due to rotation of L and T dwarfs have been reported by multiple groups  in the IRAC 4.5 $\micron$ and 8 $\micron$ bands \citep{morales06}, and at shorter near-infrared wavelengths \citep{clarke08,artigau09,radigan12}. Ongoing large Spitzer and Hubble Space Telescope programs are identifying a larger number of previously unknown varying brown dwarfs.  

\subsection{Photometric and Spectroscopic Variability of Directly-Imaged Giant Planets}
\label{Variability}

We showed that rotation periods can be measured with existing adaptive optics systems (Section~\ref{8mRegAO}) if the modulation is as prominent as our models predict ($\sim20\%$). The detection of variability with lower amplitudes will be possible with next-generation adaptive optics systems. One such system (LBT/AO), for example, already provides such a capability \citep{skemer12}. Rotational variability can provide three different types of information for
directly imaged exoplanets: 1) single-band photometry can determine the rotational period without the ambiguity of inclination; 2) multi-band photometry can constrain the
heterogeneity of the surface and the relative colors of the features, and 3) spectral
mapping can provide spectrally and spatially resolved maps of the ultra cool atmospheres, offering detailed insights into the atmospheric structures of giant exoplanets. In the following we will briefly discuss these different measurements 
and their observational requirements as predicted by our simple model.

We showed that rotation periods can already be measured with existing adaptive optics systems (Section \ref{8mRegAO}). We find that for planets broadly similar to those in the HR~8799 system the amplitude of predicted variations peaks in the J-band, yet the optimal wavelength for observations is the Ks-band, where AO performance is superior. 

Next-generation AO systems on large telescopes will provide greatly improved AO performance at shorter wavelengths. Our models predict that these systems will be able to detect rotational variations at multiple wavelengths, determining cloud covering fraction and providing color information on the surface features (see Section~\ref{8mExAO}).

We also briefly explored spectral mapping of exoplanets on the example of JWST. 
These observations will provide spatially and spectrally resolved maps over a broad
wavelength range and unaffected by telluric absorption. Our models predict that such observations both with JWST and with instruments on 30m telescopes will be limited
by instrumental stability and not by contrast. This fact highlights the importance for
sub-percent-level flux calibration techniques.   

\begin{figure}
\epsscale{1.0}
\plotone{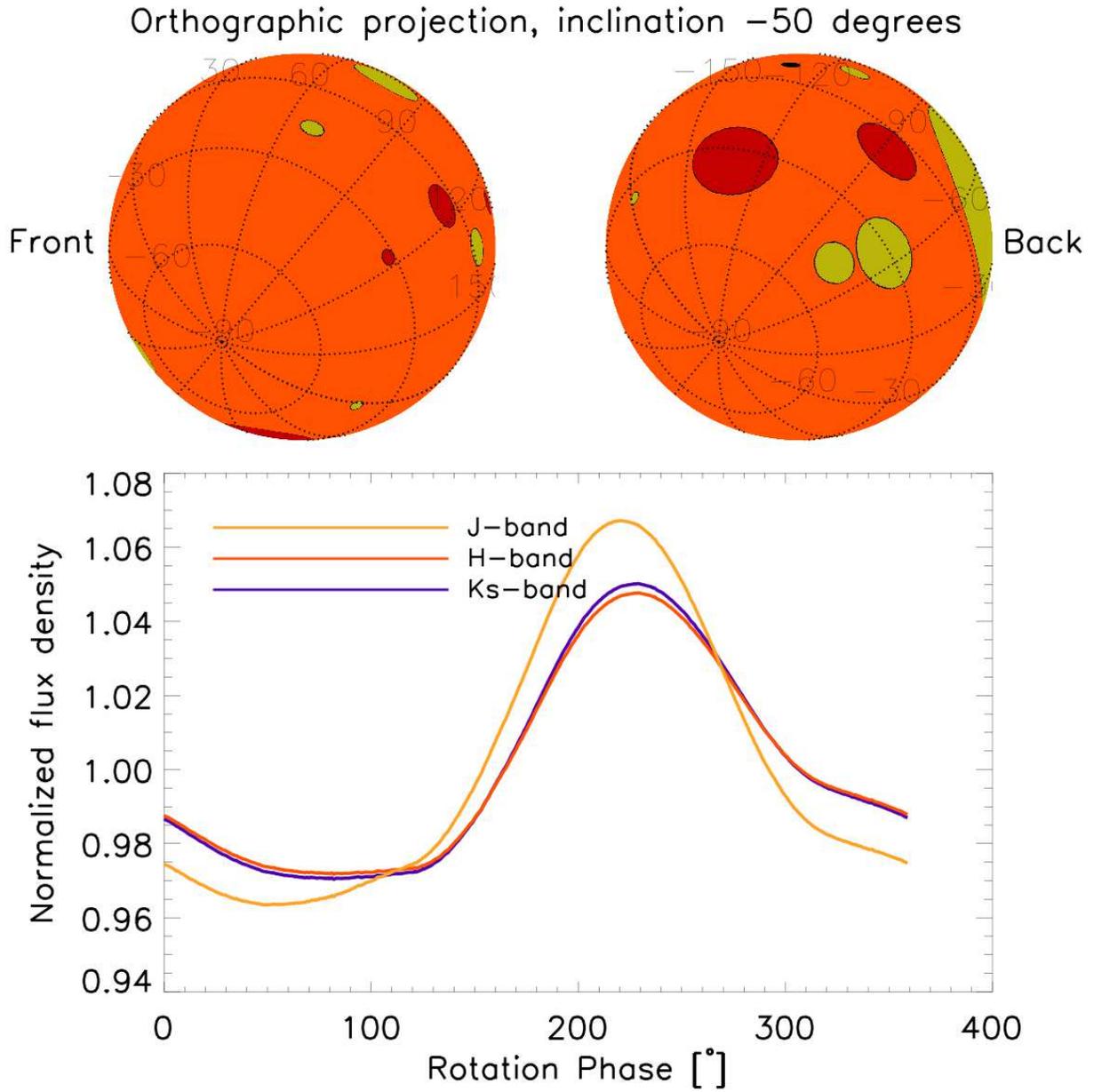}
\caption{Top panel: the same as Figure \ref{fig1} but for an inclination of $-50^\circ$. Lower panel: the same as Figure \ref{fig4} but again for an inclination of $-50^\circ$. The maximum amplitudes in all filters are significantly smaller (by as much as 12\% in J--band) compared to an inclination of $0^\circ$. The shapes of the lightcurves change as well (compared to \ref{fig4}), emphasizing the lessened contribution from the giant hot spots.\label{fig14}}
\end{figure}

We also use our models to evaluate the factors that determine the detectability of rotational variations. We find that temperature, spatial and size distributions of the spots and the effective temperature of the planet play the most important roles. For the spot distributions we modeled (Fig.~\ref{fig1}), changing the inclination to $+30^\circ$ does not affect the amplitude of the lightcurves compared to $0^\circ$ inclination but slightly changes their shape. On the contrary, an inclination of $-50^\circ$ emphasizes the contribution from the two cold spots in the right panel of Figure \ref{fig1} and changes not only the shape of the lightcurves but also the maximum amplitudes (see Figure \ref{fig14}) -- they decrease by 12\%, 9.5\% and 10\% in the J--, H-- and Ks--bands respectively compared to the maximum amplitudes for the $0^\circ$ inclination scenario seen in Figure \ref{fig4}. The importance of the inclination of the planet also depends on the spatial distribution of the spots -- if they are distributed mostly around the equatorial region, highly-inclined planets will show lower level variability than planets seen from their equatorial plane.

We note the following major difference between the Clear and Cloudy B06 models we use that have major effects on the results presented in this paper: a) The flux densities of the Cloudy (yellow, Figure \ref{fig3}) and Clear (green, Figure \ref{fig3}) B06 models of 1,000~K are similar in both H and K-bands but differ significantly in the J--band, where the Clear model spectrum is notably higher. This is the well-known J-brightening effect \citep{burgasser00,burgasser02,leggett00} caused by the appearance of cloud-free regions at the L/T transition; b) The J--band is stronger than the H--band for the 1,000~K Cloudy (yellow, Figure \ref{fig3}) but the two bands have similar strength for the 1,200~K Cloudy (light orange, Figure \ref{fig3}) models. The Ks-band is significantly weaker for both temperatures; and c) the 1,400~K Cloudy (dark orange, Figure \ref{fig3}) model spectrum peaks in the H-band and the flux densities in the J and Ks-bands are somewhat similar. Therefore, we can expect that a heterogenous surface consisting of multiple patches (spots, holes and/or clouds) described by different spectra will have significantly different observational signatures compared to a homogeneous, one-surface-type atmosphere. 

The distribution and sizes of cloud structures on directly imaged giant exoplanets will not be known a priori but constraints can be obtained by the model presented here. In Section 3 we showed that even a moderate spot total covering factor of 10\% and spot distribution similar to that of Jupiter with a dominant giant spot can produce up to 20\% photometric variability in J, H and Ks bands for a Cloudy surface with T$_{eff}$ of 800~K, 1,000~K and 1,200~K, as seen from Figure \ref{fig7}. Also, a giant spot with a temperature difference as small as $\delta T=100$~K compared to the effective temperature can cause photometric modulations of up to 20\% in Ks--band for an effective temperature of $T_{eff} = 800~$K. We also argued that the most appropriate filter, where the largest simulated photometric variability consistently occurs depends on the different surface types present on the planet. As discussed in Section \ref{Section3.1}, for our Model B2 and for the map in Figure \ref{fig1} the filter with the largest variations is Ks for T$_{eff}$ lower than 1,000~K and J for T$_{eff}$ ranging from 1,000~K to 1,400~K. However, for Model A2 the largest photometric variations occur in Ks for all T$_{eff}$ from 700K to 1,400~K. This suggests that simply detecting the wavelength at which the largest photometric variations occur will already be a strong indicator of the relative contributions of cloudy and cloud--free regions to the planet's atmosphere and also of the temperatures of these regions.

\cite{barman11a} report a weak CH$_{4}$ absorption in both H-- and K--bands and a triangular shape in the H--band spectrum for the case of HR8799b, indicating low surface gravity which may promote more efficient vertical mixing,  deviation from chemical equilibrium and indicate a young age. As noted by the authors, such low surface gravity would also imply that the condensation curve crosses the T-P curve near the photosphere and clouds can form in the deeper, photospheric depths, suggesting that the giant young planets can be cloudy and cool. For the same effective temperature and surface gravity, the thickness of the cloud layer is significantly smaller compared to the case of chemical equilibrium but still sufficient enough to produce the observed colors, much redder than those of a cloud-free brown dwarf. Cloudy, cool, low-gravity brown dwarfs have indeed been  reported \citep{stephens09} at the L/T transition, pushing the transition temperature from 1,300~K (at log(g) of 5) down to 1,100~K (at log(g) of 4.5), supporting earlier evidence for such gravity--transition regime connection \citep{metchev06,saumon08}. The rotational phase mapping proposed here can provide critical insights into the structure and distribution of cloud layers and can test cloud scale height and compositional variations in directly imaged giant planets (Section \ref{Results}).

Our model predicts that the detection of photometric and spectroscopic variations with current instrumentation are limited by a combination of contrast and stability (section \ref{Results}), but next-generation extreme AO systems will provide superior contrasts that will no longer be the limiting factor. The sensitivity limits of the current state--of--the--art instruments like VLT/NACO limit the capabilities of these instruments to measuring rotation periods. With next--generation instruments like GPI, SPHERE, or extreme AO systems on GMT, TMT or E-ELT will be able to explore the details of the cloud asymmetry through the lower--amplitude (down to an assumed photometric precision of 1\%) rotational modulations and study the J$-$Ks and H$-$Ks colors of spots covering the same fraction ($\sim 1\%$) of the surface area while JWST may push this limit down to the level of $\sim10^{-4}$. The real power of the future very large--aperture telescopes, both ground-- and space--based like GMT, TMT, E-ELT, JWST and possibly ATLAST, will be in the spectral modulation domain, where they will open up the possibility to study in detail the cloud colors, composition and/or spectra, weather patterns and even rings and satellites. 

The Great Red Spot on Jupiter has already been shown to be a very dynamic feature -- \cite{asay09} showed that the spot has shrunk by 15\% over a period of 10 years. Such behavior needs to be taken into account when studying the atmospheres of directly--imaged giant exoplanets -- variations on timescales different than the rotation period can indicate evolving atmospheric patterns or even storms. If the atmospheric patterns of these planets do indeed change on timescales shorter than the rotation period, the periodic photometric and/or spectroscopic modulations will be significantly modified or even completely erased \citep{goldman05}. Detections of photometric and/or spectroscopic evolution on timescales different from the rotation period will be a further step toward understanding the atmosphere of giant exoplanets. Such measurements, possibly achievable by JWST as discussed in Section \ref{JWST}, will require very high-cadence observations and could be an indication of differential rotation \citep{artigau09, marley10}. We have not commented on the possible presence of symmetric belts or bands in the atmosphere of these planets as they will not cause photometric or spectroscopic variability and cannot be constrained by the model presented here. 

\section{Conclusions}
\label{Conclusions}

We used simple models to explore light curve inversion as a tool to probe the surface brightness distribution and cloud properties of directly imaged exoplanets. We constructed models using state-of-the-art spectral libraries and simulated
observations with current and future instruments. The key results of this study are as follows:

i) Current AO systems on large telescopes are capable of detecting large photometric variations and thus measure
rotation periods. Cadence is limited by the slowly varying PSF systematics. 

ii) Next-generation extreme AO systems will be capable of detecting even low-level ($\sim$1\%) variations, enabling
the study of cloud heterogeneity and its wavelength dependence. These systems will also enable the study of the cloud
cover evolution.

iii) Extremely large telescopes and JWST will provide spectral mapping data for clouds, allowing detailed composition
maps of the cloud cover and the abundance of gas-phase absorbers. These setups will also provide a much higher cadence.

iv) For objects at the L/T transition, a B2 model with $T_{eff}=1000K$ and a $1200K$ hot spot predicts the largest photometric variations to be in the J--band; for cooler sources or sources
with very cool surface features the ideal wavelength gradually shifts toward 3~$\mu$m.
 
v) We demonstrated that simulated data can be inverted to a correct, low-resolution one-dimensional map of a giant
planet.

The observations proposed here will allow detailed studies of the structure and composition of condensate 
cloud covers in directly imaged exoplanets and provide otherwise inaccessible insights into the atmospheric circulation
of these exciting objects. 

\section{Acknowledgments}

A Director's Discretionary research grant by the Space Telescope Science Institute was essential for
starting this program. We are grateful to Markus Kasper for valuable discussions. We would also like like to acknowledge Justin Rogers for encouraging discussions and for his careful review of this paper. 

\bibliographystyle{apj}
\bibliography{references}

\clearpage

\clearpage
\begin{table}[ht]
\begin{center}
\begin{tabular*}{17cm}{ p{4 cm} | p{5 cm} | p{4cm} | p{4cm}}
\hline
\hline
\centering Telescope & \centering Instrument & \centering Wavelength [$\micron$] & Reference \tabularnewline
\hline
\centering 8--m class ExAO & \centering SPHERE/GPI/PISCES & \centering 1 to 2.5 & 1 \tabularnewline
\hline
\centering 30--m class ExAO & \centering GMT/TMT/ELT & \centering 1 to 5 & 2 \tabularnewline
\hline
\centering JWST & \centering NIRCAM/MIRI & \centering 1 to 27 & 3 \tabularnewline
\hline
\centering ATLAST & \centering --- & \centering 0.1 to 2.4 & 4 \tabularnewline
\hline
\hline
\end{tabular*}
\caption{List of instruments we explore: [1] \cite{mesa11,macintosh08,mccarthy01}; [2] \cite{gmt06,macintosh06,kasper08,kasper10a}; [3] \cite{stiavelli08} [4] \cite{postman10} \label{Table1}}
\end{center}
\end{table}

\clearpage
\begin{table}[ht]
\begin{center}
\footnotesize
\rotatebox{90}{
\begin{tabular*}{19cm}{ p{2.5cm} | p{1.5cm} | p{1.8cm} | p{1.8cm} | p{1.8cm} | p{2.3cm} | p{2.3cm} | p{1.5cm} }
\tableline
\hline
\centering Instrument & \centering NACO & \centering SPHERE & \centering GPI & \centering PISCES & \centering PFI & \centering EPICS & \centering NIRCAM \tabularnewline
\hline
\centering Class & 8m AO & 8m ExAO & 8m ExAO & 8m ExAO & 30m+ ExAO & 30m+ ExAO & \centering JWST\tabularnewline
\hline
\centering Filter & \centering Ks & \centering J & \centering H & \centering H & \centering H & \centering J & \centering J \tabularnewline
\hline
\centering Separation [$\arcsec$] & \centering 1 & \centering 1 & \centering 1 & \centering 1& \centering 0.2 &\centering 1 \tabularnewline
\hline
\centering 5--$\sigma$ Contrast & \centering $\sim10^{-4}$ & \centering $\sim10^{-7}$ & \centering $\sim10^{-8}$ & \centering $\sim10^{-6}$ & \centering $\sim10^{-8}$ & \centering $\sim10^{-9}$ & \centering $\sim10^{-7}$ \tabularnewline
\hline
\centering Reference & \centering 1 & \centering 2 & \centering 3 & \centering 4 & \centering 5 & \centering 6 & \centering 7 \tabularnewline
\hline
\tableline
\end{tabular*}
}
\caption{Sensitivity limits for the instruments we explore at the respective angular separation: [1] \cite{kasper07,kasper09} [2] \cite{vigan10,mesa11} [3] \cite{macintosh08} [4] \cite{mccarthy01,skemer12} [5] \cite{macintosh06} [6] \cite{kasper08,kasper10a} [7] \cite{green05} \label{table2}}
\end{center}
\end{table}

\clearpage
\begin{table}[ht]
\begin{center}
\begin{tabular*}{13.2cm}{ p{1.5cm} | p{1.5cm} | p{1.4cm} | p{1.5cm} | p{1.4cm} | p{1.5cm} | p{1.4cm} }
\hline
\hline
\centering Model & \multicolumn{2}{l|}{Effective Surface} & \multicolumn{2}{l|}{Spot Group 1} & \multicolumn{2}{l}{Spot Group 2} \tabularnewline
\cline{2-7}
 & \centering Type & \centering Temp & \centering Type & \centering Temp & \centering Type & \centering Temp \tabularnewline
\hline
\hline
\centering A1 & \centering Clear & \centering T$_{1}$ & \centering Cloudy & \centering T$_{1}$ & \centering Cloudy & \centering T$_{1}$ \tabularnewline
\hline
\hline
\centering A2 & \centering Clear & \centering T$_{1}$ & \centering Cloudy & \centering T$_{2}$ & \centering Cloudy & \centering T$_{3}$ \tabularnewline
\hline
\hline
\centering B1 & \centering Cloudy & \centering T$_{1}$ & \centering Clear & \centering T$_{1}$ & \centering Clear & \centering T$_{1}$ \tabularnewline
\hline
\hline
\centering B2 & \centering Cloudy & \centering T$_{1}$ & \centering Clear & \centering T$_{2}$ & \centering Clear & \centering T$_{3}$ \tabularnewline
\hline
\hline
\centering C & \centering Clear & \centering T$_{1}$ & \centering Clear & \centering T$_{2}$ & \centering Clear & \centering T$_{3}$ \tabularnewline
\hline
\hline
\centering D & \centering Cloudy & \centering T$_{1}$ & \centering Cloudy & \centering T$_{2}$ & \centering Cloudy & \centering T$_{3}$ \tabularnewline
\hline
\hline
\end{tabular*}
\caption{Six different realizations of the three distinct surface types (Effective Surface, Spot Group 1, Spot Group 2) covering the atmosphere of a giant planet: A1 and A2) cloudy spots on a clear surface; B1 and B2) clear spots on a cloudy surface; C) clear surface with cold and hot clear spots; and D) cloudy surface with cold and hot cloudy spots. T$_{1}$, T$_{2}$ and T$_{3}$ represent three different temperatures.\label{Table3}}
\end{center}
\end{table}

\end{document}